\documentclass[floats,aps,epsf,amsfonts]{revtex4}

\usepackage{graphicx}
\usepackage{color}
\usepackage{amsfonts,amssymb,amsmath}
\usepackage{epsfig}
\usepackage{subfigure}
\usepackage{hyperref}

\renewcommand{\b}{\bar}
\newcommand{\f}{\frac}

\newcommand{\nn}{\nonumber}

\newcommand{\el}{\ell}
\newcommand{\h}{\hat}
\renewcommand{\t}{\tilde}
\newcommand{\mrm}{\mathrm}
\newcommand{\be}{\begin{equation}}
\newcommand{\ee}{\end{equation}}
\newcommand{\ba}{\begin{eqnarray}}
\newcommand{\ea}{\end{eqnarray}}

\begin{document}

\title{A Fast Frequency-Domain Algorithm for Gravitational Self-Force:\\ Circular Orbits in Schwarzschild Spacetime}

\author{Sarp Akcay}
\affiliation{ School of Mathematics, University of Southampton,
Southampton, SO17 1BJ, United Kingdom.}
\email{sa18g09@soton.ac.uk}


\begin{abstract}
Fast, reliable orbital evolutions of compact objects around
massive black holes will be needed as input for gravitational wave
search algorithms in the data stream generated by the planned
Laser Interferometer Space Antenna (LISA). Currently, the state of
the art is a time-domain code by [Phys. Rev. D{\bf 81}, 084021,
(2010)] that computes the gravitational self-force on a
point-particle in an eccentric orbit around a Schwarzschild black
hole. Existing time-domain codes take up to a few days to compute
just one point in parameter space. In a series of articles, we
advocate the use of a frequency-domain approach to the problem of
gravitational self-force (GSF) with the ultimate goal of orbital
evolution in mind. Here, we compute the GSF for a particle in a
circular orbit in Schwarzschild spacetime. We solve the linearized
Einstein equations for the metric perturbation in Lorenz gauge.
Our frequency-domain code reproduces the time-domain results for
the GSF up to $\sim 1000$ times faster for small orbital radii. In
forthcoming companion papers, we will generalize our
frequency-domain computations of the GSF to include bound
(eccentric) orbits in Schwarzschild spacetimes, where we will employ the method of extended homogeneous solutions [Phys. Rev. D {\bf 78}, 084021 (2008)].
We will eventually extend our methods to attempt a frequency-domain computation of the GSF in Kerr spacetime.

\end{abstract}
\maketitle

\section{Introduction}
With the start of the upgrades to second generation ground based
gravitational  wave detectors \cite{Ligo,Virgo} and the approval
of the LISA Pathfinder mission \cite{Lisa_path}, the age of
gravitational wave (GW) astronomy has begun. One promising source
of gravitational radiation is the so-called extreme mass ratio
inspirals (EMRIs) where a compact object (a black hole or a
neutron star) of a few solar masses slowly spirals in toward a
massive black hole (MBH). The compact object (CO) interacts with
its own gravitational field, which causes it to move on a path
perturbed from the geodesic of the background spacetime. Along
this `forced' trajectory, the object radiates gravitationally
losing energy and angular momentum. For CO to MBH mass ratios of
$\sim 10^{-5}-10^{-6}$, the frequency of the gravitational waves
emitted during the last few years of inspiral (up to the final
plunge) will be a few mHz, which will fall right in the middle of
LISA's frequency band \cite{Lisa}. Analysis of the waveforms
emanating from these inspirals will provide us with an
unprecedented way of mapping spacetime around the central objects
\cite{Ryan}, which are presumed to be Kerr black holes. A typical
LISA bandwidth EMRI will be a $\sim 1.5 M_\odot $ neutron
star/black hole inspiraling onto a $\sim 10^6 M_\odot $ MBH. In
its last year before the plunge, the compact object will spiral in
from a distance of $\sim 10GM/c^2 $ to the innermost stable
circular orbit ($6GM/c^2 $ for Schwarzschild black hole) executing
$ \sim 5\times10^4 $ orbits and sweeping the GW frequency band
from $ \sim 2$ mHz to $\sim 5$ mHz \cite{Thorne}. Such sources
will be detectable by LISA for years, but the amplitude of the
resulting gravitational wave strain will be smaller than the noise
in the instrument \cite{Pau}. However, matched-filtering the
signal over an extended period of time ($\sim$ few years) will
bump the signal-to-noise ratio as high as $100$ for the nearest
sources \cite{BaCu}.  To be able to use matched-filtering, very
accurate gravitational wave templates will be required as input
for the cross-correlation. This will call for very accurate
simulations of these inspirals over their LISA bandwidth
lifetimes. The most challenging part in obtaining reliable
simulations will be keeping track of the orbital phase as over the
course of the inspiral the accumulated phase error should not
exceed a few radians out of a total of $ \mathcal{O}(10^5) -
\mathcal{O}(10^6) $ radians. This will put quite a stringent limit
on the error tolerance of orbital evolution models.

This is where the gravitational self-force comes in. In the test mass ($
\mu = 0$) case, the CO follows a geodesic of the background
spacetime. However, for a small, but finite mass the CO (modelled
as a point particle sourced by a Dirac delta function) interacts
with its own gravitational field, which scatters off the curvature
of the background spacetime. This interaction can be interpreted
as perturbing the particle's path off the background geodesic. In
other words, the particle now accelerates, thus feels a net force
due to this back-reaction. This is what has become known as the
gravitational self-force (GSF).

The study of radiation reaction began not with the GSF but with
electromagnetic self-force (SF). This problem was first
successfully worked out by DeWitt \& Brehme \cite{DeBr}. Later,
the solution to the gravitational problem was formulated by Mino,
Sasaki \& Tanaka \cite{MiSaTa} and independently by Quinn \& Wald
\cite{QuWa} in terms of ``forced geodesics'' where the compact
object feels a net force and is pushed off the geodesic of the
unperturbed background spacetime. This approach is generally known
as the MiSaTaQuWa formulation. Detweiler \& Whiting \cite{DeWh}
provided an alternate formulation based on geodesics of a
perturbed spacetime. These were followed by \cite{BOI, BaMiNaOSa,
Barack2001, BOII, BOIII}, which developed more practical methods
for computing the actual self-force in Schwarzschild and Kerr
spacetimes. They employed the so-called ``mode-sum scheme'' in
which the scalar, vector or the tensor perturbation is decomposed
in terms of corresponding spherical harmonics. In the case of GSF,
a tensor spherical harmonic decomposition of the retarded metric
perturbation $ \bar{h}_{\mu\nu} (t,\mathbf{x}) $ is performed.
Then, the resulting 10 second order coupled partial
differential equations are solved numerically at each tensor mode
($\el,m$). The resulting metric fields and their derivatives are
added together in certain combinations. These combinations are
then translated from tensor modes to scalar $(l,m)$ modes to yield
individual $l$ modes of the `full' GSF given by
Eq.(\ref{eq:F_full2}). As the full GSF is singular at the location
of the particle, a regularization procedure is undertaken. In
MiSaTaQuWa formulation, this is done by decomposing the divergent
`direct' part of the GSF into scalar spherical harmonics then
removing these from the full GSF at each $l$ mode. The resulting
regularized $l$ modes are finite and yield a convergent sum. This
sidesteps the issue of dealing with infinities. The final GSF is
then given by summing over the $l$ modes from zero to infinity.

The mode-sum scheme has thus far been implemented by several
groups for SF computations \cite{Burko2000} - \cite{Keidl2006}.
Most of these have been for scalar field SF or
looked at simplified cases for GSF computations (in Schwarzschild)
such as radial infall or a static particle. The GSF for circular
orbits in Schwarzschild was first successfully calculated (in time
domain) by Barack \& Sago \cite{BSI}. This was soon-after followed
by independent calculations by Detweiler \cite{Detweiler2008} and
Berndtson \cite{Berndtson2009}. Although these calculations used
different gauges and methods, by comparing the effects of the GSF
on gauge invariant quantities derived by Detweiler
\cite{Detweiler2008}, these three independent GSF computations
were shown to be equivalent \cite{Berndtson2009, BaDeSa}. The
state of the art for GSF computations is the recent work of Barack
\& Sago on eccentric orbits in Schwarzschild spacetime
\cite{BSII}. Some progress has also been made for GSF computations
in Kerr spacetime, the state of the art being the work of
Warburton \& Barack \cite{BaWar} on scalar field SF for bound
(eccentric, equatorial) orbits in Kerr spacetime. This work was
successfully implemented in frequency domain using the recently
developed method of extended homogeneous solutions \cite{BOS}.
This was a very important step in the efforts to compute the GSF
using frequency-domain methods. The method of extended homogeneous
solutions successfully avoids the `Gibbs phenomenon' that causes
the radial derivatives of the metric fields to be averaged out
across the point particle as opposed to displaying the expected
finite jump there, which is the result of modeling the particle as a
delta-function distribution. A very thorough introduction to the fundamentals
of the self-force problem is presented by Poisson
\cite{PoissonLR}. In addition, a recent article by Barack
\cite{BarackCQG} overviews the current state of the field.

Our aim in this part I of the series is to provide a fast
framework for computing the GSF that can be used for orbital
evolutions. For this reason, we have chosen to work in frequency
domain (f-domain). Berndtson \cite{Berndtson2009} was the first to
successfully compute the GSF for circular orbits in Schwarzschild
using f-domain methods, but his method differs from ours and his
work is unpublished. Starting with Regge \& Wheeler's (RW)
standard tensor harmonic decomposition of the metric perturbation
\cite{ReWh}, Berndtson solved the field equations in Lorenz gauge
by relating the gauge invariant RW, Zerilli master functions
\cite{Monc} to the unknown metric fields of Lorenz gauge. It
turned out, however, that he did not have the correct expression
for the contribution of the monopole mode to the GSF. But when he
adopted Detweiler \& Poisson's \cite{DP} solution for this mode, the results he
obtained for the GSF matched those of \cite{BSI}. His results also
highlighted the key advantages of a f-domain computation, namely,
higher accuracy and faster runtimes compared to time-domain
methods.

Despite the evident success of Berndtson's approach, it is our
feeling that our f-domain approach is better suited for extension
to Kerr in that it relies less on the spherical symmetry of the
background spacetime. As there currently exist no tensor
spheroidal harmonics, we must rely on a tensor spherical harmonic
decomposition of the metric perturbation in Kerr. The problem then
is that the resulting ordinary differential equations (ODEs)
couple between different multiple modes, not just metric fields.
However, the principal parts of the ODEs remain uncoupled and it
is possible to numerically solve the resulting system of coupled
ODEs by treating the extra couplings as new source terms. We
refrain from elaborating further as this problem is beyond the
scope of this article but our longterm research program includes
tackling these issues.

The obvious advantage of working in the f-domain is that one deals
only with ODEs, which can be solved efficiently using numerical
methods. Furthermore, in f-domain, there are no instabilities
associated with the non-radiative modes (monopole, dipole) that
one encounters in the time domain \cite{BLII, BSI}. However, there
are downsides to working in the f-domain. One is that f-domain
methods work only for bound orbits. Also, it is generally thought
that f-domain computations of GSF are intractable beyond
eccentricities of approximately $  0.7 $ \cite{GlHuKen}. The
breakdown of f-domain computations is caused by the fact that as
the eccentricity increases, there are more and more radial
frequency modes per given azimuthal mode. This significantly
augments the runtimes of numerical computations. Eventually, one
expects to reach a threshold eccentricity at which the use of
time-domain methods becomes numerically more efficient. It is
likely that f-domain methods become computationally inefficient
(compared to time-domain) at eccentricities higher than $0.7$. We
hope to empirically determine this threshold value in our future
work. However, this may not necessarily present a problem since
EMRI orbits circularize \cite{Peters} as they shrink toward the
last stable orbit and despite recent findings \cite{Pau},
\cite{BaCu} that we should expect to see EMRIs with moderate
eccentricities in the LISA bandwidth, most of the eccentricity
will have been reduced by the time the compact object begins its
final year of inspiral so that there should be plenty of EMRIs
with eccentricities $ \lesssim 0.2 $ for LISA to detect. For such
eccentricities, we expect an f-domain code to be significantly
faster than its time-domain counterparts.


As the GSF is a gauge dependent quantity (as is the orbital
radius), we must address the issue of gauge choice used in our GSF
computations. Lorenz gauge is a common choice in perturbative
studies of curved spacetimes at linear order. One is motivated by
this gauge choice because it retains the local isotropy of the
delta-function singularity used to model the compact object
\cite{BOIV}. It also casts the field equations in a fully
hyperbolic form, which is suitable for time-domain calculations.
On the other hand, the perturbed field equations are generally
more tractable in gauges like the Regge-Wheeler (RW) \cite{ReWh}
or the radiation gauges \cite{Chr}. However, thanks to the work of
Barack \& Lousto \cite{BL}, we now have access to all of the field
equations in Lorenz gauge and can follow an ``all-Lorenz-gauge''
path. This is especially desirable in the mode-sum scheme because
the multipole modes of the metric perturbation ($
\bar{h}_{\mu\nu}^{\el m}(t,r) $ ) in Lorenz gauge are continuous
($C^0$) at the location of the particle. This is not the case, for
example, in RW gauge where the source contains a
derivative-of-delta-function term in addition to the usual delta
function. Therefore, the so-called ``master functions'' used in
the RW formalism exhibit a jump-discontinuity ($C^{-1}$) at the
location of the compact object. Finally, one can compute only the
radiative ($\el \ge 2$) modes of the perturbation using approaches
based on RW gauge \cite{Ev_Hop}.

Our treatment here is mostly based on the work of Barack \& Lousto
\cite{BL} (henceforth BL) and Barack \& Sago 2007 \cite{BSI} (BS),
which use the mode-sum method in Lorenz gauge. We begin with the
linearized Einstein equations in Schwarzschild background in
Lorenz gauge. We then rewrite the field equations using tensor
spherical harmonic decomposition of the metric perturbations. This
decouples the angular part of the field equations. The resulting
set of 10 second order partial differential equations are
separated into 7 even and 3 odd parity equations. Next, we go into
the frequency domain and obtain $ 7 \oplus 3 $ second order ODEs.
For a generic bound orbit, we would need to sum over radial and
azimuthal frequency modes to work in f-domain, but for circular
orbits we have only one fundamental (azimuthal) frequency.
Therefore, the crucial step in moving to an f-domain computation
for circular orbits is supplying appropriate boundary conditions
for the metric fields. Here, we present these boundary conditions
(BC) for the first time.

With the BC specified, we numerically solve the coupled
homogeneous ODEs then impose junction conditions at
the location of the particle to construct the inhomogeneous
solutions. Once we construct all the metric perturbations and
their derivatives at the particle, we compute the GSF by using the
formulae derived in BS. This gives us what is called the ``full
self-force''. It contains a `tail' contribution, which we
interpret as the relevant physical piece and a `direct' part,
which must be removed via the appropriate regularization
procedure. It should be iterated that the initial decomposition of
the metric perturbation is done in \emph{tensor} spherical
harmonics, whereas the regularization is performed using
\emph{scalar} spherical harmonics. This requires us to translate
each tensor $(\el,m) $ mode to various scalar $(l,m)$ modes before
regularizing. This causes a single scalar mode $l $ to couple to
many tensor modes $\el$. The formulae for these couplings have
been derived by BS. Here, we use their results to compute the GSF.

For circular orbits, only the $r$-component of the GSF needs be
regularized. In the mode-sum scheme, this is done mode-by-mode at
each scalar multipole $l $ where the singular piece is decomposed
in \emph{scalar} spherical harmonics then is removed from the full
self-force at each $l$. The resulting regularized $l$ modes
have $ l^{-2} $ large-$l$ behavior, which yields a convergent
(albeit somewhat slow) sum over $l$. The physical self-force is
obtained by summing over all the individual regularized $l$ modes
and finally adding a large-$l$ ``tail'' that estimates the total
contribution due to $l > l_{max} $ modes where $l_{max} $ is the
largest mode at which we actually compute the metric
perturbations.

Section \ref{sec:intro} presents the field equations and their
decomposition under tensor spherical harmonics. In section
\ref{sec:f_dom}, we go into f-domain by Fourier transforming the
time dependence of the metric fields in azimuthal frequency modes.
We then separate the resulting field equations under their parity
and calculate the BC for each case separately. Once the BC are
known, the numerical ODE solver integrates the field equations to
yield the homogeneous solutions. Using these, we assemble the
inhomogeneous solutions, which we use in section \ref{sec:GSF} to
construct the full GSF, which we then regularize. Finally, we
compute the tail contribution to the $r$-component of the GSF. The
results are all displayed in section \ref{sec:results} where we
compare the $t$-,$r$-components of the GSF computed by our code
with that of BS. We find an excellent agreement with BS within
their error bars for orbital radii up to $\sim 100 GM/c^2$.

Throughout this article, we use geometrized units with  $ G = c= 1
$. $x^\mu= (t, r, \theta, \phi ) $ are the standard Schwarzschild
coordinates and $\tau $ denotes proper time. We follow the usual
convention of $ (-,+,+,+)$ for the metric signature. Finally,
owing to the spherical symmetry of Schwarzschild spacetime, we
work with equatorial $\theta=\pi/2$ orbits without loss of
generality.

\section{Field Equations}\label{sec:intro}
The physical set-up is that of a point particle with mass $\mu$ in
a circular orbit with radius $r_0$ around a Schwarzschild black
hole with mass $M$. The particle interacts with its own
gravitational field and thus feels a net force which moves it off
the geodesics of the background spacetime. The equation of motion
for the particle in this context is given by
\be \mu u^\mu \nabla_\mu u^\nu = F^\nu_{\mrm{GSF}} \label{eq:EOM}, \ee
where $ u^\mu \equiv dx^\mu/d\tau $ denotes the 4-velocity of the
particle, $\tau$ is proper time, $\nabla_\mu$ is the covariant
gradient operator associated with the background Schwarzschild
metric and $ F^\mu_{\mrm{GSF}} $ is the gravitational SF. Imposing
the condition that the 4-velocity remain normalized along the
worldline i.e. $ u_\mu u^\mu = -1 $ on Eq. (\ref{eq:EOM}), we get
the orthogonality condition on the self-force: $ u_\mu
F^\mu_{\mrm{GSF}} = 0 $. For circular orbits, $ F^t_{\mrm{GSF}},
F^\phi_{\mrm{GSF}} $ can be calculated independently using energy
balance arguments \cite{CKP} because they are purely dissipative.
However, in the case of eccentric orbits all non-zero components
of the SF will be made up of both dissipative and conservative
parts. The orthogonality condition is useful because it gives us a
simple way to obtain one out of the three components of the GSF
(fourth component $ F^\theta_{SF} = 0 $ because of spherical
symmetry).

To obtain the GSF in this ``forced geodesic'' picture, we must solve the perturbed Einstein's equation in a non-flat background. Schematically, the field equations have the following form
\be G_{\mu\nu} [\mathring{g}_{\mu\nu} + h_{\mu\nu} ] = 8\pi T_{\mu\nu} \label{eq:fieldEqs}, \ee
where $G_{\mu\nu} $ is the Einstein tensor, which is a functional
of the spacetime metric $g_{\mu\nu} =
\mathring{g}_{\mu\nu}+h_{\mu\nu} $ and $T_{\mu\nu}$ is the
energy-momentum tensor sourced by the point particle. Here,
$\mathring{g}_{\mu\nu} $ denotes the background (vacuum)
Schwarzschild metric and $ h_{\mu\nu}$ is the perturbation due to
the point particle. As is standard with current GSF computations,
we retain only the linear order $\mathcal{O}(\mu)$ perturbation.
There are ongoing efforts to incorporate second order
perturbations in the calculations of GSF \cite{Pound, Rosenthal},
but the current formulations are not yet ready for use in mode-sum
GSF computations.

After keeping up to $\mathcal{O}(h_{\mu\nu})$ terms in
Eq.(\ref{eq:fieldEqs}), we substitute $ G[\mathring{g}] = 0 $ into
Eq.(\ref{eq:fieldEqs}) since $\mathring{g}_{\mu\nu}$ is the metric
of a vacuum spacetime. We make two more simplifications,
which are standard: first, we change from using $h_{\mu\nu}$ to
the trace-reversed $\b{h}_{\mu\nu} $ via $\bar{h}_{\mu\nu} = h_{\mu\nu} -
\frac{1}{2}g_{\mu\nu} h $. Then, we pick a gauge. For reasons explained above and detailed in the cited articles, we choose to work in Lorenz gauge
where $ \nabla_\mu \b{h}^{\mu\nu} = 0 $. With these modifications
inserted into Eq. (\ref{eq:fieldEqs}) we obtain
\be  \Box \bar{h}_{\mu\nu} + 2 \mathring{R}_{\;\mu\,\nu}^{\alpha\,\beta} \bar{h}_{\alpha\beta} = -16 \pi T_{\mu\nu}, \label{eq:fieldEqs2} \ee
where $\Box = \nabla_\mu \nabla^\mu $.
The energy-momentum tensor is given by
\be T_{\mu\nu} = \mu \int_{-\infty}^\infty (-\mathring{g})^{-1/2} \delta^4[x^\mu-x_0^\mu(\tau)]\: u_\mu u_\nu d\tau , \label{eq:T_munu} \ee
where $x^\mu(\tau)$ denotes the position of the particle. The
proper time $\tau$ is related to the coordinate time $t$ via $
d\tau = (u^t)^{-1} dt $. Finally, $\mathring{g}$ is the
determinant of the Schwarzschild metric equaling $ -r_0^4 $ for
$\theta = \pi/2$.

As it stands, Eq. (\ref{eq:fieldEqs2}) represents 10 coupled
$2^{\mrm{nd}}$ order, partial differential equations (PDEs). We
can simplify these by separating out the angular part. To this
end, we decompose $\b{h}_{\mu\nu}(t,\mathbf{r})$ using
\emph{tensor} spherical harmonics, which form a 10-dimensional
basis for any rank two, symmetric 4-dimensional tensor field. The
components of the metric perturbation are decomposed as follows:
\be \bar{h}_{\mu\nu}(t,\mathbf{r}) = \frac{\mu}{r}\sum_{\ell, m}
\sum_{i=1}^{10}  \bar{h}^{(i)\ell m}(t,r) Y^{(i)\ell
m}_{\mu\nu}(\theta,\phi; r) \label{eq:Ten_harm} . \ee
The explicit expressions for $Y^{(i)\el m}_{\mu\nu} $ are
presented in BL. We modify them slightly here: $Y^{(i)\el
m}_{\mu\nu\:\mrm{here}} = a^{(i)\el} Y^{(i)\el m}_{\mu\nu\:
\mrm{BL}}  $ where $a^{(i)\el} $ constant coefficients defined in
BL.
Now angular variables decouple and the field equations become (at each $\el,m$)
  \begin{equation}
  \Box_{sc} \bar{h}^{(i)\el m} + \mathcal{M}^{(i)}_{(j)} \bar{h}^{(j)\el m} = \mathcal{S}^{(i)\el m}  .\label{eq:fieldEqs3}
 \end{equation}
where $ f = f(r) \equiv 1-\f{2M}{r} $ and $ \Box_{sc} $ is the usual scalar field wave operator:
\be \Box_{sc} = \frac{1}{4}\left[\partial_t^2 - \partial^2_{r_\ast} + f \left(\frac{2M}{r^3} + \frac{\ell (\ell+1)}{r^2}\right) \right]. \label{eq:box_sc} \ee
$\mathcal{S}^{(i)}$ are the source terms obtained from decomposing $T_{\mu\nu}$ in tensor spherical harmonics. They are given by
\be \mathcal{S}^{(i)\el m}=4\pi \tilde{E}_0 \: \alpha^{(i)} \times
\delta(r-r_0) \left\{\begin{array}{ll} [Y^{\el m }(\theta,
\Omega_0 t)]^\ast & i=1,2,\ldots, 7 \\ \partial_\theta[ Y^{\el
m}(\theta, \Omega_0 t)]^\ast & i=8,9,10 \end{array}\right.
\label{eq:S_i} \ee
where $ \tilde{E}_0 = (1-2M/r_0)/\sqrt{1-3M/r_0} $ is the
dimensionless energy of a test particle ($\mu = 0$) on a circular
geodesic with radius $r_0$. Given the orbital angular frequency
$\Omega_0 = d\phi_0/dt =  (M/ r_0^3)^{1/2}$, the constants
$\alpha^{(i)} $ are:
\ba
\alpha^{(1)} & = & f^2_0/r_0, \quad \alpha^{(3)}  = f_0/r_0, \qquad \alpha^{(7)} = r_0 \Omega_0^2 [\el(\el+1)-2m^2], \nn \\
\alpha^{(2),(5),(9)} &=& 0 , \qquad \qquad \alpha^{(8)} = 2 f_0 \Omega_0, \nn \\
\alpha^{(4)} &=&2 i f_0  m \Omega_0, \qquad \alpha^{(10)} = 2 i m r_0 \Omega_0^2, \nn \\
\alpha^{(6)} &=& r_0 \Omega_0^2 ,\label{eq:alphas}
\ea
where $f_0= 1-2M/r_0$. Note that the $ (i) = 2,5,9 $ equations are sourceless. The spherical harmonics are given by the usual formula
\be Y^{\el m }(\theta,\phi) = \hat{c}_{\el m} P^{\el m}(\theta) e^{i m \phi} \label{eq:Y_lm} . \ee
$P^{\el m}(\theta) $ are the associated Legendre polynomials and $ \hat{c}_{\el m} \equiv \sqrt{\f{2\el+1}{4\pi}\f{(\el-m)!}{(\el+m)!}} $. We can further rewrite the second line in Eq. (\ref{eq:S_i}) using the following expression:
\ba \partial_\theta [Y^{\el m}(\pi/2, \phi_0)]^\ast &=& \left[\el C_{\el+1,m} \hat{c}_{\el+1,m}\: P^{\el+1,m}(\pi/2)-(\el+1) C_{\el m} \hat{c}_{\el m} \:P^{\el-1,m}(\pi/2) \right] e^{-i m \phi_0} \nn \\ & \equiv & \mathcal{J}^{\mrm{(odd)}}_{\el m} e^{-i m \phi_0} \label{eq:J_odd}
\ea
where $ C_{\el m} = \sqrt{\f{\el^2-m^2}{(2\el+1)(2\el-1)}} $.

The $\mathcal{M}^{(i)}_{(j)} \b{h}^{(j)} $ in Eq.
(\ref{eq:fieldEqs3}) contain the coupling terms between different
field equations. In the next section, we will show that up to 5
field equations couple together for certain modes, but things will
not get any more entwined than that. The expressions for
$\mathcal{M}^{(i)}_{(j)} \b{h}^{(j)} $ are lengthy and have been
given in detail in \cite{BSII}, \cite{BL} and \cite{BSI} so we
omit them here. We will however present the field equations in
frequency domain in section \ref{sec:f_dom}.

Eq. (\ref{eq:S_i}) substituted in to Eq. (\ref{eq:fieldEqs3}) gives us the Einstein field equations in their simplest form that we can reach in Lorenz gauge. From this point, one can either go into time domain and tackle the problem of solving these coupled PDEs or one can go into frequency domain and deal with ODEs that require boundary conditions. In the next section, we solve the field equations in frequency domain in Lorenz gauge for the first time.

\section{Frequency-Domain Solutions of The Field Equations}\label{sec:f_dom}
Here, we begin by decomposing the metric fields $\b{h}^{(i)}(t,r)
$ into frequency modes. In the case of circular orbits, there is
only one frequency: $ \Omega_0 = (M/r_0^3)^{1/2} $. So the
harmonics of circular motion are given by $\omega_m = m \Omega_0
$. For elliptical orbits, the frequency modes will be a
combination of azimuthal and radial fundamental frequencies: $
\omega_{m n} = m \Omega_\phi + n \Omega_r $. For circular orbits,
metric fields are decomposed as follows:
\be \bar{h}^{(i)\el m}(t,r) =  R^{(i)}_{\el m}(r) e^{-i \omega_m
t}, \qquad \mathrm{where} \ \omega_m = m \Omega_0 . \ee
This reduces the 2-dimensional hyperbolic equations
(\ref{eq:fieldEqs3}) to a set of $2^{\mrm{nd}}$ order, coupled
ODEs, which can be numerically solved much more quickly than PDEs
encountered in time-domain approaches. In the case of a scalar
field in Schwarzschild spacetime, the problem in f-domain reduces
to a single inhomogeneous ODE. The standard procedure is to
numerically solve for the \emph{homogeneous} inner ($r<r_0$) and
outer ($r>r_0$) solutions then construct the inhomogeneous
solution by imposing the correct junction conditions at $r=r_0$.
For the computation of the GSF, the same procedure applies but now
for many coupled fields, some of which have delta-function sources
and others no sources at all. In section \ref{sec:Inhomog}, we
explicitly show how we construct the inhomogeneous solutions from
coupled homogeneous solutions.

The system of 10 coupled, second order homogeneous ODEs can be
written as
\be \f{d^2 R^{(i)}_{\el m}(r)}{d r^2_\ast} - 4 V_{\el
m}(r)R^{(i)}_{\el m}(r) - 4 \tilde{\mathcal{M}}^{(i)}_{(j)}
R^{(j)}_{\el m}(r) = 0 \label{eq:R_field_eq} . \ee where $ r_\ast
$ is the Regge-Wheeler tortoise coordinate with $ dr_\ast/dr =
f^{-1} $, $\tilde{\mathcal{M}}_{(j)}^{(i)}$ is the Fourier
transformed version of $\mathcal{M}_{(j)}^{(i)}$, and
\be V_{\el m}(r) = \f{1}{4} \left[\f{2M f}{r^3} + \f{\el(\el+1)f}{r^2} - \omega_m^2 \right] \label{eq:V_eff} . \ee
The field equations (\ref{eq:R_field_eq}) are not all coupled to
each other; our 10-dimensional basis splits under parity very much
like in Regge-Wheeler gauge. The $ (i)=1, \ldots, 7 $ basis
elements of the tensor spherical harmonics are \emph{even} and the
$(i)=8, 9, 10 $ basis elements are \emph{odd} under parity
transformations. For circular orbits, \emph{even, odd} mean that
$\el+m = \mrm{even},\: \mrm{odd} $. Eqs. (\ref{eq:R_field_eq}) now
decouple completely under these two parity sectors so they can be
solved completely independently. Furthermore, because the
spherical harmonics in the source terms (\ref{eq:S_i}) give $
[Y^{\el m}(\pi/2, \phi_0)]^\ast = 0 $ for $\el+m=\mrm{odd} $ and $
\partial_\theta [Y^{\el m}(\pi/2,\phi_0) ]^\ast = 0 $ for
$\el+m=\mrm{even} $, the odd parity solutions are trivially zero
for an even mode and vice versa for even parity solutions. That is
$ R^{(1) \ldots (7)} = 0$ for $\el+m= \mrm{odd}$ and
$R^{(8),(9),(10)}=0$ for $\el+m =\mrm{even}$.

Similarly, the four gauge equations coming from the Lorenz gauge condition $\nabla_\mu \bar{h}^\mu_{\ \nu} = 0 $ also decouple under parity with three equations falling under the even parity sector, leaving only one for the odd sector. The gauge equations at each ($\el,m $)-mode are
\be
 i \omega_m R^{(1)} + f \left( i \omega_m R^{(3)} + R^{(2)}_{,r} + \f{R^{(2)}}{r} - \f{R^{(4)}}{r} \right) = 0, \label{eq:gauge1}
\ee
\be
-i \omega_m R^{(2)} - f R^{(1)}_{,r} + f^2 R^{(3)}_{,r} -\f{f}{r} \left( R^{(1)} - R^{(5)} - f R^{(3)} - 2f R^{(6)} \right) = 0, \label{eq:gauge2}
\ee
\be
-i \omega_m R^{(4)} - \f{f}{r} \left( r R^{(5)}_{,r} + 2 R^{(5)} + \el (\el+1) R^{(6)} - R^{(7)} \right) = 0, \label{eq:gauge3}
\ee
\be
-i \omega_m R^{(8)} - \f{f}{r} \left(r R^{(9)}_{,r} + 2 R^{(9)} - R^{(10)} \right) = 0 . \label{eq:gauge4}
\ee
Here and henceforth, we omit writing the modal indices $\el,m $ as
well as the functional dependence on $r_\ast$ (or $r$) for the
sake of brevity. It should be assumed that each field equation
presented holds for a given $\el,m$ mode unless stated otherwise.

Thanks to the gauge equations, it turns out that not all the even
(or odd) equations need to be solved simultaneously. As we have
four gauge conditions, we have only $10-4=6$ degrees of freedom
(d.o.f). These split as $ 4+2 $ under parity. But because of the
particular form of the field equations in the even sector, we must
solve 5 coupled ODEs together, construct the inhomogeneous
solutions then use two gauge equations to obtain the fields
$R^{(2)}$ and $R^{(4)}$ (more on this later in section
\ref{sec:Even_Sec}). In the odd sector, we solve the two coupled
$(i)=9,10$ equations together then use the odd gauge equation to
obtain $R^{(8)}$. This procedure of solving the equations in
stages is called ``the hierarchical solving scheme'' by BL. It
involves first numerically solving only the ODEs that couple to
each other then using gauge equations (\ref{eq:gauge1}) -
(\ref{eq:gauge4}) to determine the remaining unknown radial
fields. The number of equations one has to solve changes depending
on the values of $\el$ and $m$. For a generic even mode ($\el \ge
2, m > 1 $), one solves 5 coupled ODEs then uses two gauge
equations whereas for a generic odd mode ($\el \ge 2, m \ge 1$),
only two coupled ODEs are solved numerically then one gauge
equation is used. There are also non-generic modes such as the
monopole ($\el=0$); the even, odd dipoles ($\el=1, m=1,0$) and the
static ($m=0$) even, odd modes. Analytic solutions have been
explicitly provided in \cite{DP} for the monopole, and by BL for
the odd static modes. The even dipole ($\el=1,m=1$) and the $
\el=\mrm{even} $ static modes are solved numerically, but have
fewer number of non-zero fields. We present all the
different cases for both even and odd parity sectors and the
hierarchical scheme for solving the field equations in table
\ref{fig:table1}

\begin{center}
\begin{table}
\begin{tabular} {c | c | c}
    \hline \hline
    & Even ($\el+m=2N$) & Odd ($ \el+m=2N +1 $) \\
    \hline \hline
    $\el=0 $ &  $    (i) = 1, 3, 6 \rightarrow 2 $ (A)& no field  \\
    $ \el = 1 $ & $ m=1: (i)=1, 3, 5, 6 \rightarrow 2, 4 $ & $ m=0: (i)=8$ only (A) \\
    $ \el \ge 2 $ & $ (i)=1, 3, 5, 6, 7 \rightarrow 2, 4 $ &    $(i)=9,10 \rightarrow 8 $ \\
    \ & $ m=0: (i)=1, 3, 5 \rightarrow 6,7 $ & $ m=0: (i)=8$ only (A) \\
    \hline \hline
    \end{tabular}
\caption{The hierarchical solving scheme for the ten field equations. The arrows $\rightarrow$ indicate that we use the gauge equations to obtain the field to the right of the arrow. (A) indicates that the solutions are obtained analytically and $N \in \mathbb{N}$} \label{fig:table1}
\end{table}
\end{center}
\subsection{Odd Sector}\label{sec:odd_sec}
We begin with what we call {\it generic} odd modes ($m>0$). We
will consider the static odd modes ($m=0$) later in a special
subsection. As explained in the hierarchical scheme, here we solve
the coupled $(i)=9,10 $ equations together to determine $ R^{(9)}$
and $R^{(10)} $ then use these solutions in the odd gauge equation
(\ref{eq:gauge4}) to solve for $ R^{(8)} $. The two homogeneous,
odd parity field equations are \ba
\partial^2_{r_\ast} R^{(9)}& =& 4 \left[ V_{\el m}+ \f{f}{r^2}\left(1-\f{4.5M}{r}\right)\right] R^{(9)} - \f{2f}{r^2}\left(1-\f{3M}{r}\right) R^{(10)}, \label{eq:ODE_9} \\  \partial^2_{r_\ast} R^{(10)}& =& 4 \left( V_{\el m} -\f{f}{2r^2}\right) R^{(10)} - \f{2f \lambda}{r^2} R^{(9)},  \label{eq:ODE_10}
\ea
where $ \lambda = ( \el+2 )( \el-1 ) $. In order to get the
correct numerical solutions to Eqs. (\ref{eq:ODE_9}) and
(\ref{eq:ODE_10}), we must specify appropriate boundary conditions
for the numerical ODE integrator. The boundaries are located on
the event horizon ($r=2M$) and at radial infinity ($r=\infty$),
which translate to $r_\ast=-\infty$ and $r_\ast =\infty $,
respectively.  A quick inspection of the structure of the ODEs
$(i)=9, 10 $ reveals that as $r,r_\ast \rightarrow \infty  $ and $
r\rightarrow 2M (r_\ast \rightarrow -\infty)$, the $ \omega_m^2 $
term dominates in the potential and the ODEs (\ref{eq:ODE_9}) and
(\ref{eq:ODE_10}) asymptotically turn into standard wave
equations. Thus, for the solutions at infinity and on the event
horizon, we have the usual outgoing and ingoing wave behavior,
respectively. Denoting the outgoing/ingoing \emph{homogeneous}
solutions by $ R^+_i $ and $ R^-_i$, respectively, we write the
following ansatz for the boundary conditions:
\ba
R^+_{9,10} & = & e^{i \omega_m r_\ast} \sum_{k=0}^\infty \f{a^k_{9,10}}{r^k}, \label{eq:R9_BC}\\
R^-_{9,10} & = & e^{-i \omega_m r_\ast} \sum_{k=0}^\infty b^k_{9,10} (r-2M)^k \label{eq:R10_BC} .
\ea
Clearly at $r=2M$ and $ r=\infty$ we get the proper wave-like behavior. We must also specify $ dR^\pm_i/dr $  ($i=9,10$) at the boundary points. Our numerical code uses $r_\ast $ as the integration variable so we actually need $ dR^\pm_i/dr_\ast = f dR_i^\pm/dr  $ for the BC.

Numerically, we can not use infinities for the boundary points. For
our code, we pick a range of $ r_\ast \in [-65M, -55M] $ for the
inner boundary. $r_\ast=-65M$, which corresponds to $ r/M \approx
(2+10^{-14}) $ is about as far `in' as we can go due to double
floating point machine accuracy. The choice for the outer boundary
point $r_{\mrm{out}}$ depends on $\el$ and $\omega_m$ as we demand
that the outer boundary be located in the wave zone, which
translates to $ r_{\mrm{out}} \gg (\el r_0)/\omega_m $. So we opt
for an adaptive outer boundary at each ($\el,m$) where $
r_{\mrm{out}} = 50\: (\el r_0)/\omega_m $. The ratio of $50$ was
chosen after numerical experimentation. Larger ratios mean larger
runtimes for the computation of the homogeneous fields, and
smaller ratios call for more terms in the series in
Eqs.(\ref{eq:R9_BC}), (\ref{eq:R10_BC}) for numerical convergence.

Note that the sums for the BC in Eqs.(\ref{eq:R9_BC}), (\ref{eq:R10_BC}) are infinite. However, because we solve the coupled field equations numerically, we must truncate the sums at some $k=k_{\mrm{max}} $. We numerically determine this $k_{\mrm{max}} $ for each of the sums at every ($\el,m$) such that the next term in the summation has absolute magnitude less than $10^{-14} $. We also numerically check that each sum converges.

The coefficients $a_k^i$ and $b_k^i $ are unknown and must be determined by substituting our ansatz into the field equations then constructing recursion relations for the $k^{\mrm{th}}$ coefficients $a_k^i$ and $b_k^i $ out of $a_{k'< k}^i$ and $b_{k' < k}^i $. The recursion relations for the outer BC for $R^+_9$ and $R^+_{10}$ are as follows:
%
\ba
2 i \omega k\; a^9_k & =& C_{k-1}\: a^9_{k-1} + D_{k-2}\: a^9_{k-2} + E_{k-3}\: a^9_{k-3} + 2 a^{10}_{k-1} -10M a^{10}_{k-2} + 12M^2 a^{10}_{k-3} ,\nn \\ \label{eq:a9_k} \\
2 i \omega k\; a^{10}_k & = & I_{k-1}\: a^{10}_{k-1} + J_{k-2}\: a^{10}_{k-2} + K_{k-3}\: a^{10}_{k-3} + 2\lambda a^9_{k-1} - 4M\lambda a^9_{k-2} , \label{eq:a10_k}
\ea
where
\ba
C_k & = & 4M i\omega k + k(k+1) - L - 4, \qquad I_k = 4M i\omega k + k(k+1) - L + 2, \nn \\
D_k & = & -6Mk - 4M k^2 + 24M + 2ML, \quad J_k = -6Mk - 4M k^2 -6M + 2ML, \nn \\
E_k & = & 4 M^2 (k^2 + 2k-8), \qquad\qquad K_k = 4M^2 (k^2 +2k +1) . \nn
\ea
Here and in all other recursion relations that we present,
$\omega$ denotes $\omega_m = m\Omega_\phi$ and $L \equiv
\el(\el+1) $. The recursion relations are rather
cumbersome, which is why we will refrain from presenting the rest
of them in the main body of the paper unless we refer to them
directly (as done in section \ref{sec:l_even_m_zero}). All the
recursion relations are listed in appendix \ref{sec:recursion}.

The recursion relations must be started off by specifying the
values for the leading terms. In the case of odd parity equations,
these first terms are $ a^{9,10}_0 $ and $b^{9,10}_0 $ with $
a^{9,10}_{k<0} = 0 $ and $b^{9,10}_{k<0} = 0 $. This gives us 4
free parameters to specify every time we wish to solve the system
of coupled ODEs. Since we have one gauge equation and three field
equations for both inner and outer homogeneous solutions, we end
up with $2\times(3-1)=4 $ degrees of freedom. These d.o.f. are
manifest in our freedom for choosing the values for $ a^{9,10}_0 $
and $b^{9,10}_0 $. In the next subsection, we show how to pick
suitable values for these coefficients and construct the
inhomogeneous solutions.

The final remark concerns the nature of the BC specified above. As
can be clearly seen, the ingoing/outgoing wave conditions for the
BC yield complex numbers. Therefore, we must construct complex
solutions for the homogeneous fields $ R^\pm_i $. A quick
inspection reveals that the real and imaginary part of the complex
fields $ R^{(i)}$ completely decouple in the field equations
(\ref{eq:ODE_9}) and (\ref{eq:ODE_10}). As a result, we simply
solve each given ODE twice: once with the real part of the BC and
a 2$^{\mrm{nd}}$ time using the imaginary part of the BC. We then
combine the two numerical homogeneous solutions under one complex
solution that we also call $ R^\pm_i $. Recall that we already
have to solve the homogeneous ODEs twice to get the inner ($-$)
and outer ($+$) solutions and now twice more for the real and
imaginary parts. In total, at each generic odd mode, we must
numerically solve the system of coupled ODEs $2\times 4=8 $ times.

\subsubsection{Obtaining The Inhomogeneous Solutions} \label{sec:Inhomog}
To obtain the true, inhomogeneous solutions --- which are sourced
by $\delta$-functions --- we must impose junction conditions on
the coupled homogeneous solutions. Recalling that the
inhomogeneous solutions $R^{(i)}$ must be $C^0$ fields, the two
conditions are continuity at $r_0$ and the correct jump of $
dR^{(i)}/dr$ across $r_0$. Because we have coupled fields, we must
construct the inhomogeneous solutions from linear combinations of
homogeneous solutions. We use standard methods of constructing a
linearly independent basis of homogeneous solutions and imposing
the correct junction conditions to assemble the inhomogeneous
fields. Below, we briefly outline this procedure.

As mentioned before, in the odd sector we have a total of 4 d.o.f.
so we construct a 4-dimensional basis from the homogeneous
solutions $ R^\pm_9$ and $ R^\pm_{10} $. We do this by exploiting
the freedom we have in choosing the initial values for the coefficients $
a^{9,10}_{k=0}, b^{9,10}_{k=0} $ that start the recursion
relations (\ref{eq:a9_k}) - (\ref{eq:a10_k}). A linearly
independent 4-dimensional basis can be constructed for the
homogeneous solutions $ R_{9,10}$ by setting $ (a^9_0,
a^{10}_0)=(1,0) \ \mrm{ then } \ (0,1) $ and the same for $
(b^9_0, b^{10}_0)$. These determine our basis vectors at the point
of interest, namely $r=r_0$. We label these solutions by $
R^{[1]\pm}_9, R^{[1]\pm}_{10} $ and $R^{[2]\pm}_9, R^{[2]\pm}_{10}
$. For example, $R^{[1]+}_9, R^{[1]+}_{10} $ are obtained by
setting $a^9_0=1$ and $a^{10}_0=0$ then solving the coupled ODEs
(\ref{eq:ODE_9}) and (\ref{eq:ODE_10}) for $R^+_9(r_0)$ and
$R^+_{10}(r_0) $. Recall that since the boundary conditions are
complex, the basis vectors are complex as well. Finally, we follow
the same procedure for the $r$-derivatives. We label the inner and
outer basis elements for the $r$-derivatives  $ \partial_r
R^{[1]\pm }_9, \partial_r R^{[1]\pm }_{10} $ and $ \partial_r
R^{[2]\pm}_9, \partial_r R^{[2]\pm }_{10} $. So in this notation,
$ \partial_r R^{[2]-}_{10}$ stands for $ dR^-_{10}/dr|_{r_0} $
obtained by setting $ b^9_0= 0 $ and $ b^{10}_0 = 1 $.

We label the inhomogeneous solutions by $R^{(i)}_{\mrm{in}},
R^{(i)}_{\mrm{out}} $. These are constructed from
$R^{[j]+}_{9,10}, R^{[j]-}_{9,10}$ respectively. The inhomogeneous
solutions are obtained by imposing the standard junction
conditions: (1) Continuity at $r_0$: $R^{(i)}_{\mrm{in}}(r_0) =
R^{(i)}_{\mrm{out}}(r_0)  $, (2) The following jump for the
$r$-derivatives at $r_0$:
\be \left.\f{dR^{(i)}_{\mrm{out}}}{dr}\right|_{r_0} - \left.\f{dR^{(i)}_{\mrm{in}}}{dr}\right|_{r_0} = -\f{16 \pi \mu \tilde{E}_0 \alpha^{(i)}}{f^2_0} \times \mathcal{J}^{\mrm{odd}} \equiv J^{(i)}, \qquad (i) = 9, 10 \ , \label{eq:odd_jump} \ee
where $\mathcal{J}^{\mrm{odd}}$ is given by Eq. (\ref{eq:J_odd}).
To impose these conditions for our basis of homogeneous solutions,
we form a $4\times4 $ complex matrix containing the fields $
R^{[j]\pm}_i, \partial_r R^{[j]\pm}_i $ listed above. The
inhomogeneous solutions $R_{\mrm{in,out}}^{(9),(10)}$ are
constructed from linear combinations of the homogenous solutions
multiplied by unknown complex coefficients $ x_j $. To determine
these coefficients, we must solve the following matrix equation:
\be
\left(\begin{array}{cccc} -R^{[1]-}_9 & -R^{[2]-}_9 & R^{[1]+}_9 & R^{[2]+}_9 \\   -R^{[1]-}_{10} & -R^{[2]-}_{10} & R^{[1]+}_{10} & R^{[2]+}_{10} \\ -\partial_r R^{[1]- }_9 & - \partial_r R^{[2]-}_9 & \partial_r R^{[1]+}_9 & \partial_r R^{[2]+}_9 \\   -\partial_r R^{[1]-}_{10} & -\partial_r R^{[2]- }_{10} & \partial_r R^{[1]+}_{10} & \partial_r R^{[2]+}_{10}\\ \end{array} \right) \left(\begin{array}{c} x_1 \\ x_2 \\ x_3 \\ x_4 \\ \end{array} \right) = \left(\begin{array}{c} 0 \\ 0 \\ 0 \\ J^{(10)} \\ \end{array} \right)\label{eq:odd_matrix}
\ee
The right hand side (RHS) of Eq.(\ref{eq:odd_matrix}) ensures the
continuity of the inhomogeneous solutions and imposes the correct
jump value $J^{(i)} $ on the first derivatives. Recall that
because $\alpha^{(9)} = 0 $ (see Eq.(\ref{eq:alphas})), we have
$J^{(9)} =0$. We solve for the complex $ x_1, \ldots, x_4 $ by
using standard numerical matrix inversion algorithms. Once we know
the $x_1, \ldots, x_4$, we construct the inhomogeneous solutions
at the location of the particle. These are given by
\ba
R^{(i)}_{\mrm{in}}(r_0) &=& x_1 R^{[1]-}_i + x_2 R^{[2]-}_i = x_3 R^{[1]+}_i + x_4 R^{[2]+}_i = R^{(i)}_{\mrm{out}}(r_0), \label{eq:R9}  \\
\left.\f{dR^{(i)}_{\mrm{in}}}{dr}\right|_{r_0} & = & x_1
\partial_r R^{[1]-}_i + x_2 \partial_r R^{[1]-}_i, \quad
\left.\f{dR^{(i)}_{\mrm{out}}}{dr}\right|_{r_0}  =  x_3
\partial_r R^{[1]+}_i + x_4 \partial_r R^{[2]+}_i, \label{eq:dR9}
\ea
where $i=9, 10$. Although the continuity of $ R^{(9)}, R^{(10)}$
and $ dR^{(9)}/dr $ (because $J^{(9)}=0$) is analytically exact,
because the coupled ODEs are solved numerically, we will
inevitably have a small violation of continuity at $r=r_0$.  This
is caused by the numerical matrix inversion. Usually, the
numerical inversion algorithms are very robust and the
discontinuity in the fields is $ \sim 10^{-13} - 10^{-14} $ for
most modes. However, for a few special modes, this error becomes
much more significant. We will comment more on this issue later in
section \ref{sec:errors}.

We take the solutions (\ref{eq:R9}) and substitute them into the odd gauge equation (\ref{eq:gauge4}) to solve for $R^{(8)}(r_0)$.
After this step, we obtain $dR^{(8)}/dr $ at $r=r_0$ by differentiating the gauge equation  (\ref{eq:gauge4}) with respect to $r$ and using the field equation (\ref{eq:ODE_9}) to substitute for $ \partial_r^2 R^{(9)}$ term in $dR^{(8)}/dr $.
%
%
Recall that $ R^{(8)}$ has a non-zero $\delta$-function source
thus it exhibits the standard jump discontinuity at $r_0$ given by
Eq.(\ref{eq:odd_jump}). Therefore, we must compute
$dR^{(8)}/dr|_{r_0} $ twice: once as $ r\rightarrow r_0^+ $ then
again for $ r \rightarrow r_0^- $. Since $ R^{(8)}(r_0)$ and its
$\pm$ $r$-derivatives are obtained algebraically from
Eq.(\ref{eq:gauge4}) --- by inserting the numerical solutions $
R^{(9),(10)}(r_0),\: dR^{(9),(10)}/dr|_{r_0}$ --- we expect the
error in the continuity of $R^{(8)}(r_0)$ to be comparable to
errors found for $ R^{(9), (10)}(r_0) $. Indeed, we find that the
offset in the continuity of $R^{(8)}(r_0)$ is $ \sim 10^{-13} $.
Similarly, the relative error between $J^{(8)}$ and the jump of
$dR^{(8)}/dr|_{r_0}$ is $\sim 10^{-14} $.

As mentioned above, we have to solve the set of coupled ODEs $8 $
times for each odd parity mode: twice owing to the fact the BC are
complex, and 4 times because we construct the inhomogeneous
solutions from a 4-dimensional basis of homogeneous solutions.
Doing a run up to e.g. $\el_{max}=18 $, we end up with $81$
generic odd modes, which yield a total of $81\times 8 = 648$ times
that the coupled set of odd ODEs must be solved numerically.

\subsubsection{The Static ($m=0$) Odd Modes} \label{sec:odd_static_modes}
As shown in BL, the $m=0$ odd modes have analytic solutions. Since
$ J^{(10)} \propto m = 0 $ and $ J^{(9)} = 0 $, we trivially have
that $ R^{(9)}=R^{(10)} = 0 $ for these modes. Therefore, we solve
a single ODE for $ R^{(8)}$. For the case of $\el = 1 $, the ODE
simplifies to a well known form, which has the following analytic
solution:
\be
R^{(8)}_{\el=1}(r) = -\f{1}{3} r_0 \beta_{\el=1} \times \begin{cases} (r/r_0)^2, & r \le r_0 \\ (r_0/r), & r \ge r_0 , \\ \end{cases} \label{eq:R8_odd_dipole}
\ee
where $\beta_{\el=1} = 16 \sqrt{3\pi} f_0^{-1} \tilde{E}_0
\Omega_0 $ . For $ \el > 1 $, the inner ($ r<r_0 $) homogeneous
solutions exhibit the standard power law behavior: $ \sim
r^{\el+1} $. As for the outer solutions ($ r>r_0 $), we have
something that is of the form $ r^{-\el} (1 + \ln f) $. These
scale as $ r^{-\el} $ as $ r\rightarrow \infty $, which is
regular. The details of how these analytic solutions are
constructed are given in section IIIC of BL, which is why we
refrain from elaborating more here. We also omit the explicit
expressions for these  static, $\el > 1$ solutions in this
article. The interested reader should peruse BL (\cite{BL}). In
summary, the overall static, odd solutions are given by ---
restoring the modal indices --- $  \b{h}^{(9)\el 0} =
\b{h}^{(10)\el 0} = 0$ and the non-zero fields $ \b{h}^{(8)\el 0}$, which are constructed analytically .
\subsection{Even Sector} \label{sec:Even_Sec}
For the generic, non-static case of even modes, we have 7 field
and 3 gauge equations thus a total of $2\times(7-3)=8 $ d.o.f.
However, an inspection of the even parity field equations as they
are written in Lorenz gauge (\cite{BSII}, \cite{BL}, \cite{BSI})
reveals that we must simultaneously solve 5, not 4, coupled ODEs.
As before, we numerically solve the homogeneous ODEs then obtain
the inhomogeneous solutions by employing the standard techniques
for coupled fields, which we illustrated in section
\ref{sec:Inhomog}. The 5 homogeneous coupled ODEs in the even
sector are the $ (i)=1,3,5,6, 7 $ equations written in the
following form:
\begin{small}
\ba
\partial^2_{r_\ast} R^{(1)} & = & 4 V_{\el m} R^{(1)} + \f{4M}{r^2} f R^{(3)}_{,r_\ast} + \f{2f}{r^2}\left(1-\f{4M}{r}\right) \left(R^{(1)}-R^{(5)} - f R^{(3)} \right) - \f{2f^2}{r^2} \left(1-\f{6M}{r}\right) R^{(6)}, \nn \\ \label{eq:eq_R1} \\
\partial^2_{r_\ast} R^{(3)} & = & 4 V_{\el m} R^{(3)} - \f{2f}{r^2} \left[R^{(1)} - R^{(5)} - \left(1-\f{4M}{r}\right) \left(R^{(3)} + R^{(6)}\right) \right], \label{eq:eq_R3} \\
\partial^2_{r_\ast} R^{(5)} & = & 4 V_{\el m} R^{(5)} + \f{4f}{r^2} \left[ \left(1-\f{4.5M}{r}\right) R^{(5)} - \f{L}{2}\left(R^{(1)} - f R^{(3)} \right) + \f{1}{2} \left(1-\f{3M}{r}\right) \left( L R^{(6)} - R^{(7)} \right) \right], \nn \\ \label{eq:eq_R5} \\
\partial^2_{r_\ast} R^{(6)} & = & 4 V_{\el m} R^{(6)} - \f{2f}{r^2} \left[R^{(1)} - R^{(5)} - \left(1-\f{4M}{r}\right) \left(R^{(3)} + R^{(6)}\right) \right], \label{eq:eq_R6} \\
\partial^2_{r_\ast} R^{(7)} & = & 4 V_{\el m} R^{(7)} - \f{2f}{r^2} \left(R^{(7)} + \lambda R^{(5)} \right). \label{eq:eq_R7}
\ea
\end{small}
In this article, we follow the convention of BS \cite{BSI} for the
field $\b{h}^{(3)}$, which is different from that of BL \cite{BL}:
$ \b{h}^{(3)}_{here} = \b{h}^{(3)}_{\mrm{BL}}/f $. Recall that $ f
= 1-2M/r $, $ L \equiv \el(\el+1) $, $ \lambda = (\el+2)(\el-1) $
and $V_{\el m}$ is given by Eq. (\ref{eq:V_eff}). Next, we must
specify the boundary conditions. As was the case with the odd
sector fields, we impose the same ingoing/outgoing wave conditions
on the event horizon and at radial infinity, respectively. We once
again use  $R^-_{i}, R^+_{i} $ to denote the ingoing, outgoing
homogeneous solutions, respectively. For the inner/outer BC, we
use the same ansatz as before
\ba
R^-_i & = & e^{-i \omega_m r_\ast} \sum_{k=0}^\infty b_k^i (r-2M)^k , \label{eq:evenBC_Iansatz} \\
R^+_i & = & e^{i \omega_m r_\ast} \sum_{k=0}^\infty \f{a_k^i}{r^k} \label{eq:evenBC_Oansatz}  \ea
for $ i = 1, 3, 5, 6, 7 $. Once again, we substitute these ansatz into the field equations (\ref{eq:eq_R1}) - (\ref{eq:eq_R7}) to derive new recursion relations for the coefficients $ a^i_k, b^i_k $ in Eqs. (\ref{eq:evenBC_Iansatz}), (\ref{eq:evenBC_Oansatz}). The sums are of course infinite but we truncate them at some $ k = k_\mrm{{max}} $ as we did before. The recursion relations for the outer coefficients $ a_k^i $ and inner coefficients $ b_k^i $ are given in appendix \ref{sec:recursion}.

With the coefficients $a_k^i, b_k^i $ determined, there still remains one critical issue that pertains to the total number of degrees of freedom to use: in the even sector, we have 5 ODEs that can not be decoupled from each other, so we must solve all five simultaneously, but we have $ 8 $ d.o.f in the even sector, not $2\times 5= 10$. So, there must be an extra condition on each set of 5 BC for inner and outer homogeneous solutions. For the outer solutions, this extra condition is a constraint on the coefficients $ a^3_k $, which is given by the even gauge equations:
\be a^3_0 = 0 \label{eq:gauge_a3}. \ee
We repeat this procedure of eliminating the $5^{\mrm{th}}$ degree
of freedom from the inner homogeneous solutions by making use of
the gauge equations. After some manipulation, we reach the
following condition on the coefficients $ b^k_3 $:
\be
b^3_0 = -\f{\left[ \left( \: i\el(\el+1) + 4 M\omega(1- 4 M i \omega + \el(\el+1)\: )\: \right) b^1_0 + i\: (1 + 16 M^2 \omega^2 ) b^5_0 \right]}{2 M \omega(1+16M^2 \omega^2)} \label{eq:gauge_b3} .
\ee
So all of the coefficients $ b^3_k $ are entirely determined from $ b^1_0, b^5_0 $ and the recursion relation (\ref{eq:inBC_3}). With the conditions (\ref{eq:gauge_a3}) and (\ref{eq:gauge_b3}) imposed, we are left with the expected 8 d.o.f.

Eqs.(\ref{eq:gauge_a3}) and (\ref{eq:gauge_b3}) tell us that our 8-dimensional basis of inner and outer homogeneous solutions is constructed by using the recursion relations (\ref{eq:outBC1}) - (\ref{eq:inBC_7}) for the BC with $ \{b^1_0, b^5_0, b^6_0, b^7_0\} $ and $ \{ a^1_0, a^5_0, a^6_0, a^7_0\} $ as the sets containing the 8 free parameters for the inner and outer homogeneous solutions. We construct our basis of linearly independent homogeneous solutions by numerically determining the basis vectors that span the solution space. Each basis vector of the outer homogeneous solution space is obtained by setting one of the coefficients $ \{ a^1_0, a^5_0, a^6_0, a^7_0\} $ equal to $1$ while the other $3$ equal 0. We do this a total of 4 times, e.g.  $ \{a^1_0, a^5_0, a^6_0, a^7_0\} = \{ 1,0,0,0\}, \{ 0,1,0,0\},\{ 0,0,1,0\} $ and $\{ 0,0,0,1\} $. This procedure is repeated with $ \{b^1_0, b^5_0, b^6_0, b^7_0\} $ for the inner solutions. This yields 8 basis vectors for constructing the 8-dimensional linearly independent homogeneous solution space. Given that the system of ODEs must be solved twice because the BC are complex, we reach a total of $ 8 \times 2 = 16 $ for the number of times we must numerically solve the field equations at each \emph{even} mode. For example, for $\el$ running up to $18$, we have a total of $89$ generic even modes, which means that the coupled ODEs are numerically integrated a total of $89\times 16 = 1424$ times. This is what takes up the main bulk of our numerical computation time. We will say more about this later. Next, we construct the inhomogeneous solutions.

\subsubsection{Inhomogeneous Solutions}\label{sec:even_inhomo}
In subsection \ref{sec:Inhomog}, we showed in detail how to
construct the inhomogeneous solutions from the inner and outer
homogeneous solutions. Here we do the same with the even parity
solutions. Our basis of homogeneous solutions is now 8-dimensional
and is spanned by $ R^\pm_i, \partial_r R^\pm_i $ with $ i=1,5,6,7
$. In accordance with the notation of subsection
\ref{sec:Inhomog}, we label the basis vectors (the homogeneous
fields $ R^\pm_{1,5,6,7} $) by $ R^{[j]\pm}_i $. Similarly, for
the derivatives, we use $
\partial_r R^{[j]\pm}_i   $. For example, $R^{[1]+}_1 $ stands for
$R_1^+$ obtained by setting $a^1_0 = 1$ and $ a^5_0 = a^6_0 =
a^7_0 = 0 $ and $ \partial_r R^{[3]-}_6 $ is $ dR^-_6/dr $ with
$b^6_0 = 1 $ and $ b^1_0= b^5_0 = b^7_0 = 0 $.

To construct the inhomogeneous solutions, we impose the junction
conditions on the homogeneous fields and their $r$-derivatives in
the form of an 8-dimensional complex matrix equation:
\be \left( \begin{array}{c|c} -R^{[j]-}_i  &  R^{[j]+}_i  \\ \hline  -\partial_r R^{[j]- }_i  &  \partial_r R^{[j]+}_i \end{array} \right) \left( \begin{array}
{c} x_1 \\ \vdots \\ x_8 \end{array}\right) = \left( \begin{array}
{c}  0_{4\times 1} \\  J^{(i)} \end{array} \right) . \label{eq:8by8matrix} \ee
$ 0_{4\times 1} $ is a $4 \times 1$ array of zeros imposing the condition of continuity for the inhomogeneous fields $R^{(i)}$ and
\be J^{(i)} \equiv -\f{16\pi \mu \t{E}_0 \alpha^{(i)}}{f_0^2} \hat{c}_{\el m} P^{\el m}(\theta=\pi/2) . \label{eq:J_even2}\ee
%
The complex, inhomogeneous fields $ R^{(i)}$ at $r=r_0$ are given by
\be
R_{\mrm{in}}^{(i)}(r_0) = \sum_{j=1}^4 x_j R^{[j]-}_i = \sum_{j=1}^4 x_{j+4} R^{[j]+}_i = R_{\mrm{out}}^{(i)}(r_0) . \label{eq:Ri_even} \\
\ee
Similarly, for the $r$-derivatives of these fields at $r=r_0$, we have
\ba
\left.\f{dR^{(i)}_{\mrm{in}}}{dr}\right|_{r_0} & = & \sum_{j=1}^4\:x_j \partial_r R^{[j]-}_i  \label{eq:dRi_in} \\
\left.\f{dR^{(i)}_{\mrm{out}}}{dr}\right|_{r_0} & = & \sum_{j=1}^4
\: x_{j+4} \partial_r R^{[j]+}_i. \label{eq:dRi_out} \ea
%
%
We still need to determine the inhomogeneous field $R^{(3)}$ and
its $r$-derivative at $r_0$. Recall that in order to form the
linearly independent basis of homogeneous solutions we had to
solve a system of 5 (not 4) coupled ODEs together. However, the
homogeneous solutions $ R^\pm_3 $ and their first derivatives $
R'^\pm_3 $ are not part of our basis because they are constructed
from linear combinations of the other basis elements as shown in
Eqs. (\ref{eq:gauge_a3}) \& (\ref{eq:gauge_b3}). With the basis of
homogenous solutions at hand, $R^{(3)}(r_0), dR^{(3)}/dr|_{r_0} $
are simply given by
\ba
R^{(3)}_{\mrm{out}}(r_0) & = & \sum_{j=1}^4 x_j R^{[j]-}_3 = \sum_{j=1}^4 x_{j+4} R^{[j]+}_3    = R^{(3)}_{\mrm{out}}(r_0), \label{eq:R3_even}\\
\left.\f{dR^{(3)}_{\mrm{in}}}{dr}\right|_{r_0} & = & \sum_{j=1}^4 \:x_j \partial_r R^{[j]-}_3  \label{eq:dR3_in}, \\
\left.\f{dR^{(3)}_{\mrm{out}}}{dr}\right|_{r_0} & = & \sum_{j=1}^4
\:x_{j+4} \partial_r R^{[j]+}_3. \label{eq:dR3_out} \ea
The remaining two fields $ R^{(2)}$ and $ R^{(4)} $ are extracted from the even parity gauge equations (\ref{eq:gauge2}), (\ref{eq:gauge3}). Their $r$-derivatives are obtained by differentiating these gauge equations with respect to $r$ and substituting the relevant parts of the fields equations $(i)=1,3,5$ for the $ \partial_r^2 R^{(1)}, \partial_r^2 R^{(3)}, \partial_r^2 R^{(5)}$ terms that arise from $r$-derivatives of Eqs. (\ref{eq:gauge2}), (\ref{eq:gauge3}).

Although $R^{(i)}_{\mrm{in}}(r_0) = R^{(i)}_{\mrm{out}}(r_0)$
analytically, because we invert the complex matrix numerically, we
are bound to have small discontinuities at $r_0 $ as we did with
the odd parity fields. We checked the relative error in the
continuity of the fields $R^{(1)}, \ldots, R^{(7)}$ at $r=r_0$ and
found that it is at most $ \mathcal{O}(10^{-12}) $ for $r_0
\lesssim 100M $ and $\el-m = \: \mrm{small}$. However, we find
that for $r_0 > 100M $, as $\el-m \rightarrow 15 $, the violation
of the continuity of the field $R^{(5)}$ grows up to
$\mathcal{O}(10^{-7}) $ in relative size. For $\el-m \gtrsim 30 $,
this violation climbs up to $ \mathcal{O}(10^{-5}) $.
Clearly, for large orbital radii and large $\el-m$, the numerical matrix inversion becomes less accurate. A quick check of condition numbers $c$ for the matrices in Eq. (\ref{eq:8by8matrix}) shows that $ c \gtrsim 10^{12} $ for the problematic cases mentioned here. We explain the cause of this in section \ref{sec:errors}. However, it is only the field $R^{(5)}$ that exhibits the bad discontinuities; the fields $ R^{(1),(6),(7)} $, which also come directly out of the matrix inversion, have continuity violations that are consistently at least three or more orders of magnitude smaller. As expected, larger inversion errors persist in the fields $R^{(2)}, R^{(4)}$ (and their $r$-derivatives) because these are constructed from gauge equations containing $R^{(5)}$ and its first and second $r$-derivatives. As far as we can tell this matrix inversion error, which we quantify by the numerical discontinuity of the fields $R^{(2),(4),(5)}$ at $r=r_0$ is our largest source of error. We will say more on this inversion error in section \ref{sec:errors}. 
%
%
%
%
\subsubsection{The Even Dipole ($\el=1, m=1 $) Mode}
The even parity dipole mode is non-radiative ($\el < 2 $) thus
represents a shift in the orbital angular momentum, which can be
interpreted as a rotation of spacetime around its center of mass.
For $\el=1,m=1 $, $ \lambda = 0 $ as well as $ \alpha^{(7)} =
J^{(7)} = 0 $. This gives $ \bar{h}^{(7)\:11}(t,r) = 0 $, which
results in 4 coupled ODEs. The $(i)=1,3,6 $ equations
(\ref{eq:eq_R1}), (\ref{eq:eq_R3}), (\ref{eq:eq_R6}) do not
contain any $ R^{(7)} $ terms as such they remain unchanged, as do
the recursion relations for $a^{1,3,6}_k, b^{1,3,6}_k$ displayed
in appendix \ref{sec:generic_rec}. However the $(i)=5 $ equation
(\ref{eq:eq_R5}) does contain  a $ \lambda R^{(7)} $ term, which
is now zero so we end up with new recursion relations for the
inner and outer boundary conditions for $R^\pm_5 $. These are
given by Eqs. (\ref{eq:inBC_5_1_1_A}), (\ref{eq:outBC5_1_1_A}) in
appendix \ref{sec:even_dipole_BC}.

With $R^\pm_7=0$, we have $2\times(6-3)=6$ degrees of freedom for
our basis of homogeneous solutions. The basis vectors are
constructed from the homogeneous solutions obtained by using the
BC generated from the sets $\{b^1_0, b^5_0, b^6_0\} $ and $
\{a^1_0, a^5_0, a^6_0\} $, respectively. The ODE integrator solves
the coupled system a total of $2\times 6 =12 $ times. To obtain
the inhomogeneous solutions, we construct a $6\times 6 $ complex
matrix very similar to the one in Eq.(\ref{eq:8by8matrix}), but
without the homogeneous fields $ R^{[j]\pm}_7, \partial_r
R^{[j]\pm}_7 $. We solve the resulting matrix equation to obtain
the values for the complex amplitudes $ x_1, \ldots, x_6 $, which
in turn, give us the values of the inhomogeneous solutions and
their first $r$-derivatives at $r_0 $. The equations for the
inhomogeneous fields $ R^{(1),(3),(5),(6)}(r_0) $ are identical to
Eq. (\ref{eq:Ri_even}), (\ref{eq:R3_even}) with $x_7=x_8=0 $. The
fields $ R^{(2),(4)}(r_0) $ are once again obtained from the gauge
equations (\ref{eq:gauge2}) and (\ref{eq:gauge3}) with $R^{(7)} =
0 $.

\subsubsection{The Monopole $\el=0 $ Mode} \label{sssec:mono}
This conservative, non-radiative $\el=0$ contribution to the
metric perturbations represents a shift in the mass of the small
particle across $r=r_0$. For this mode, the field equations
simplify enough that analytic solutions have been found by
Detweiler \& Poisson \cite{DP}. The only non-zero
fields are $ \b{h}^{(1)}=R^{(1)},\ \b{h}^{(3)} = R^{(3)}, \
\b{h}^{(6)} = R^{(6)} $, which contribute only to the diagonal
(scalar) components of $ h_{\mu\nu} $. In section III.D of BL, the
solutions are displayed explicitly in terms of the components of $
h_{\mu\nu} $. As with the other modes, these are $ C^0 $ with the
usual jump in the $r$-derivative across $r_0 $. We omit writing
the explicit solutions here and refer the interested reader to
\cite{BL}, section III.D for the details. The extra important step
we mention here is the rewriting of these analytic solutions ---
written as components of $h_{\mu\nu}$ in BL --- in terms of
$\b{h}^{(i)}$. Although this seems like a backward step, it is
necessary in order to properly follow the algorithm for computing
the GSF. We will elaborate more on this procedure later in section
\ref{sec:GSF}.

The formulae needed to transform $ h_{tt}, h_{rr}, h_{\theta\theta}$, $ h_{\phi \phi} $ to $ \b{h}^{(1)}, \b{h}^{(3)}, \b{h}^{(6)} $ are as follows (\cite{BL}):
\ba
\b{h}^{(1)}_{\el=0}(r) & = & 2 \sqrt{\pi} \mu^{-1} r \left(h_{tt} + f^2 h_{rr} \right) \label{eq:h1_mono}, \\
\b{h}^{(6)}_{\el=0}(r) & = & 2 \sqrt{\pi} \mu^{-1} \f{r}{f} \left(h_{tt} - f^2 h_{rr} \right) \label{eq:h6_mono}, \\
\b{h}^{(3)}_{\el=0}(r) & = & 4 \sqrt{\pi} \mu^{-1} r^{-1} h_{\theta\theta} = 4 \sqrt{\pi} \mu^{-1} r^{-1}(\sin\theta)^{-2} h_{\phi \phi}. \label{eq:h3_mono}
\ea
Note that the expression for $\b{h}^{(3)} $ here looks different
from the one given by BL in \cite{BL}. The reader may recall that
this is because we use the $\b{h}^{(3)} $ as defined by BS in
\cite{BSI} as opposed to BL as was mentioned earlier . From these
relations and the explicit expressions provided for
$\bar{h}^{(1),(3),(6)}$ in \cite{BL}, it is straightforward to evaluate
the fields $\bar{h}^{(i)}$ and their inner and outer $r$-derivatives at
$r=r_0 $, which then give us the total contribution of the
monopole ($\el=0$) to the GSF.

\subsubsection{The Even Static Modes ($\el\ge 2 (\mrm{ even}), m=0 $)}\label{sec:l_even_m_zero}
These modes require a special discussion not only because the
dimension of the homogeneous solutions space is smaller but also
because the BC require extra care. With $m=0 $, we have that $
\alpha^{(2)} = 0 $ and $ \alpha^{(4)} = 0 $ . Furthermore, an
inspection of $ tr, t\theta, t\phi $ components of $h_{\mu\nu} $
(cf. Eq.(20) of \cite{BL}) reveals that these depend only on
$\bar{h}^{(2)} $ and $ \bar{h}^{(4)} $. Since static modes must be
symmetric under time reversal, we have that $ h_{ti} = 0 $ for $
i=r,\theta,\phi $ thus we must have $\bar{h}^{(2)} = 0$ and
$\bar{h}^{(4)} = 0 $ for the static, even modes. This reduces the
total number of fields in the even sector to 5 and eliminates the
gauge equation (\ref{eq:gauge1}) (it gives the trivial $ 0 = 0 $).
Using the remaining two gauge equations (\ref{eq:gauge2}),
(\ref{eq:gauge3}), we can obtain expressions for $R^{(6),(7)}$ in
terms of $R^{(1),(3),(5)}$. We then substitute these into the
field equations (\ref{eq:eq_R1}) - (\ref{eq:eq_R5}). This yields
modified field equations for $(i)=1,3,5 $:
\begin{small}
\ba
\partial^2_{r_\ast} R^{(1)} & = & 4 V_{\el m} R^{(1)} + \f{4M}{r^2} f \partial_{r_\ast} R^{(3)} + \f{2f}{r^2}\left(1-\f{4M}{r}\right) \left(R^{(1)}-R^{(5)} - f R^{(3)} \right) \nn \\ &\quad & - \f{f}{r^2} \left(1-\f{6M}{r}\right) \left[R^{(1)} + \f{r}{f} \partial_{r_\ast} R^{(1)} - f R^{(3)} - r \partial_{r_\ast} R^{(3)} - R^{(5)} \right], \label{eq:eq_R1_stat} \\
\partial^2_{r_\ast} R^{(3)} & = & 4 V_{\el m} R^{(3)} \label{eq:eq_R3_stat} \\ & - & \f{2f}{r^2} \left\{R^{(1)} - R^{(5)}  - \left(1-\f{4M}{r}\right) \left[R^{(3)}  + \f{1}{2f}\left( R^{(1)} + \f{r}{f} \partial_{r_\ast} R^{(1)} - f R^{(3)} - r \partial_{r_\ast} R^{(3)} - R^{(5)} \right)\right] \right\}, \nn \\
\partial^2_{r_\ast} R^{(5)} & = & 4 V_{\el m} R^{(5)}  \label{eq:eq_R5_stat} \\ & +  & \f{4f}{r^2} \left[ \left(1-\f{4.5M}{r}\right) R^{(5)} - \f{\el(\el+1)}{2}\left(R^{(1)} - f R^{(3)} \right) - \f{1}{2} \left(1-\f{3M}{r}\right) \left(2 R^{(5)} + \f{r}{f} \partial_{r_\ast} R^{(5)}   \right) \right].
\nn \ea
\end{small}
Next, we calculate the boundary conditions for the static homogeneous solutions $R^\pm_{1,3,5}$. Because we are looking at static modes, the ingoing/outgoing wave conditions are no longer appropriate for the BC. Our determining criterion is now regularity, so for the inner homogeneous solutions $R^-_i $, we select the following ansatz:
\be R^-_i = \sum_{k=k_{\mrm{start}}}^\infty b^i_k (r-2M)^k \label{eq:IBC_even_stat} . \ee
Substituting the ansatz (\ref{eq:IBC_even_stat}) into the field
equations for $ (i)=1,3,5 $ gives us new recursion relations for
the BC, which we display explicitly below as we will be making
remarks about them here. We also list them in appendix
\ref{sec:even_stat_rec}.
\ba
8M^3 k (k-2) b^1_ k & = & \b{F}^1_{k-1}b^1_{k-1} + \b{G}^1_{k-2} b^1_{k-2} + \b{G}^3_{k-2} b^3_{k-2} - 2M b^5_{k-2} \nn \\
& \quad & + \b{E}^3_{k-3} b^1_{k-3} + \b{E}^1_{k-3}b^3_{k-3} - b^5_{k-3},  \label{eq:rec_b1_stat} \\
4M k (k-1) b^5_k & = & \b{C}^5_{k-1} b^5_{k-1} - 4ML b^1_{k-1} + \b{D}^5_{k-2} b^5_{k-2} + 2L (b^3_{k-2} - b^1_{k-2} ),\nn \\ \label{eq:rec_b5_stat}
\b{C}^3_{k-1} b^3_{k-1} & = & \b{C}^1_{k-1} b^1_{k-1} - 8 M^3 k b^1_k  + 4M^2 b^5_{k-1} + \b{D}^3_{k-2} b^3_{k-2} + \b{D}^1_{k-2} b^1_{k-2} \nn \\
& \quad & + 4M b^5_{k-2} + \b{E}^3_{k-3} b^3_{k-3} + \b{E}^1_{k-3} b^1_{k-3} + b^5_{k-3} , \label{eq:rec_b3_stat}
\ea
where
\ba
\b{C}^1_k & = & -4M^2 (k+1), \quad \b{C}^3_k  =  4M^2 k (k-1),\quad \b{C}^5_k = 2ML-4M(1+k^2) \label{eq:stat_C} \\
\b{D}^1_k & = & 2M(k-2), \quad \b{D}^3_k = 2M (L + k (1-2k)), \quad \b{D}^5_k = L-k(k+1), \label{eq:stat_D} \\
\b{E}^1_k &=& L +1 - k^2, \quad \b{E}^3_k = k-1, \quad \b{G}^3_k =  2Mk, \label{eq:stat_E_G}  \\
\b{F}^1_k & = & 4 M^2 (L+1+ 4k - 3k^2  ), \quad \b{G}^1_k = 2M (2L + 2 + 2k - 3k^2 ). \label{eq:stat_F_G}
\ea
Little care is needed when evaluating the coefficients $ b^1_k,
b^3_k $ using the recursion relations (\ref{eq:rec_b1_stat}) and
(\ref{eq:rec_b3_stat}). First, because the left-hand-side of
Eq.(\ref{eq:rec_b1_stat}) gives zero for $k=0,2 $ we must start
this recursion relation at $k=3 $ with $b^1_0 = b^1_1 = 0 $.
Similarly, the recursion relation for $b^5_k $ starts at $k=2 $
with $ b^5_0 =0 $ and $b^5_1 $ as the free parameter. Further
inspection reveals that the remaining two free parameters are $
b^3_0, b^3_1 $. This can be seen by realizing that $ \b{C}^3_k = 0
$ for $k=1$ so we can not use the recursion relation
(\ref{eq:rec_b3_stat}) until $k=2$ but we need $b^3_0$ and $ b^3_1
$ to determine $b^1_{k \ge 2} $ in Eq.(\ref{eq:rec_b1_stat}) and $
b^5_{k \ge 2} $ in Eq.(\ref{eq:rec_b5_stat}). So our 3-dimensional
basis of homogeneous solutions is generated from the set $
\{b^3_0, b^3_1,b^5_1 \} $. When we evaluate these three recursion
relations to obtain the higher-k coefficients, we first get $
b^1_k, b^5_k $ then at the $(k+1)^{\mrm{th}}$ order we recover
$b^3_k $. For example, at $k=2$ we obtain $ b^1_2, b^5_2 $ then at
$k=3$ we recover $ b^3_2 $ and also obtain $ b^1_3, b^5_3 $. As
usual, we truncate the infinite sum at some $k=k_{max} $ such that
the contribution of $(k_{max}+1)^{\mrm{th}} $ term has absolute
magnitude less than $10^{-14} $.

Next, we turn to determining the outer boundary conditions. This
particular case is more involved than all the other BC thus far
mentioned. First of all, the naive ansatz of $ R^+_i = \sum_k
a^i_k/ r^k $ only provides two free parameters thus falls one
short of the needed three d.o.f. for the outer solutions. Inspired
by the analytic, outer homogeneous solutions for
$\el=\mrm{odd},m=0$ modes, which have $ r^{-\el}, r^{-\el}\ln{r} $
large-$r$ behavior, we make the following ansatz
\be
R^+_i = \sum_{k=k_{\mrm{start}}}^\infty \f{a^i_k + \b{a}^i_k \ln{r}}{r^k} \label{eq:outBC_stat_even}.
\ee
When we substitute this ansatz into the ODEs (\ref{eq:eq_R1_stat}), (\ref{eq:eq_R3_stat}), (\ref{eq:eq_R5_stat}), we find that $ a^i_k = 0 $ for all $ k < \el $. Two of the three free parameters are $ a^3_{k=\el}, a^5_{k=\el} $ which combine to give
\be a^1_{\el} = a^3_{\el} + \f{a^5_{\el}}{\el+1} . \label{eq:a_k_stat_even} \ee
The next order terms in the recursion relations are as follows
\ba a^1_{\el+1} & = & \f{1}{4L} \left[2L (2+\el) a^1_\el- 4L a^3_\el + 2(2-\el^2) a^5_\el \right], \nn \\
a^3_{\el+1} & = & \f{1}{4L} \left[ 2L (\el-2) a^1_\el + 12L a^3_\el + 2( \el(\el+2)-2) a^5_\el \right], \nn \\
a^5_{\el+1} & = & \el a^3_{\el+1} + (\el+2) a^1_{\el+1} - 2(\el^2-\el-2) a^3_\el - 4 a^1_\el . \nn
\ea
Note that all of these still only depend on the 2 free parameters $ a^3_{\el}, a^5_{\el} $. It turns out the third free parameter is $ a^5_{\el+2} $. As for the $\b{a}^i_k $, they are all given in terms of $ \{a^3_\el, a^5_\el, a^5_{\el+2}\} $ with the condition $ \b{a}^i_{k < \el+2} = 0 $. Unlike the previous cases, here we get two sets of recursion relations from each field equation, one for $ a^i_k $ and another for $\b{a}^i_k $. These are:
\ba
\h{C}^1_k a^1_k&=& (k+1) a^3_k + a^5_k - 2k \b{a}^1_k - \b{a}^3_k  \nn\\
& &-2M \left(\h{D}^1_{k-1} a^1_{k-1} + \h{D}^3_{k-1} a^3_{k-1} + a^5_{k-1} + \h{E}^1_{k-1} \b{a}^1_{k-1} - 2\b{a}^3_{k-1}\right) \nn \\
& & + 4M^2 \left(\h{F}^3_{k-2} a^3_{k-2} -\b{a}^3_{k-2} \right) \label{eq:outBC1_even_stat} , \\
\h{C}^1_k \b{a}^1_k & =&  (k+1) \b{a}^3_k + \b{a}^5_k -2M \left(\h{D}^1_{k-1} \b{a}^1_{k-1} + \h{D}^3_{k-1} \b{a}^3_{k-1} + \b{a}^5_{k-1} \right) \nn \\
& & +4M^2 \h{F}^3_{k-2} \b{a}^3_{k-2} , \label{eq:outBC_bar1_even_stat}
\ea
where
\ba
\h{C}^1_k & = & L+1-k^2, \quad \h{D}^1_k = k(k-1), \quad \h{D}^3_k = 2(k+1), \nn \\
\h{E}^1_k & = & 1-2k, \quad \h{F}^3_k = k+1 . \nn
\ea
\ba
\h{C}^1_k a^3_k & = & (k+1) a^1_k - a^5_k -\b{a}^1_k - 2 k \b{a}^3_k \nn \\
& &-2M \left( \h{G}^3_{k-1} a^3_{k-1} + \h{G}^1_{k-1} a^1_{k-1} + \h{H}^3_{k-1} \b{a}^3_{k-1} - 2 \b{a}^1_{k-1} \right) \nn \\
& & + 4M^2 \left( \h{I}^3_{k-2} a^3_{k-2} + \h{J}^3_{k-2} \b{a}^3_{k-2} \right) \label{eq:outBC3_even_stat} , \\
\h{C}^1_k \b{a}^3_k & = & (k+1) \b{a}^1_k -\b{a}^5_k - 2M \left( \h{G}^3_{k-1} \b{a}^3_{k-1} + \h{G}^1_{k-1} \b{a}^1_{k-1} \right) \nn \\
& & + 4M^2 \h{I}^3_{k-2} \b{a}^3_{k-2} \label{eq:outBC_bar3_even_stat} ,
\ea
where
\ba
\h{G}^3_k & = & 2k^2-2-L, \quad \h{G}^1_k = 2k, \quad \h{H}^3_k = -4k, \nn \\
\h{I}^3_k & = & k^2-1, \quad \h{J}^3_k = -2k . \nn
\ea
\ba
\h{C}^5_k a^5_k & = & 2L (a^1_k - a^3_k) - \h{D}^5_k \b{a}^5_k + 2M \left( \h{E}^5_{k-1} a^5_{k-1} + 2L a^3_{k-1} + \h{D}^5_{k-1} \b{a}^5_{k-1} \right) \label{eq:outBC5_even_stat} ,\\
\h{C}^5_k \b{a}^5_k & = & 2L (\b{a}^1_k -\b{a}^3_k ) + 2M \left(\h{E}^5_{k-1} \b{a}^5_{k-1} + 2L \b{a}^3_{k-1} \right), \label{eq:outBC_bar5_even_stat}
\ea
where
\be \h{C}^5_k  =  L + k(1-k), \quad \h{D}^5_k = 2k-1, \quad
\h{E}^5_k  =  k(1-k) + 2 . \nn \ee
The careful reader will note that the recursion relations appear coupled to each other in Eqs. (\ref{eq:outBC1_even_stat}) - (\ref{eq:outBC_bar5_even_stat}). That is, unlike all other recursion relations, the right-hand-sides of Eqs. (\ref{eq:outBC1_even_stat}) - (\ref{eq:outBC_bar5_even_stat}) contain $k^{\mrm{th}}$ order terms. If we move all order $k$ terms to the left-hand-sides of Eqs. (\ref{eq:outBC1_even_stat}) - (\ref{eq:outBC_bar5_even_stat}), we find that the LHSs form a coupled system of 6 equations with 6 unknowns. These equations are `uncoupled' by using standard linear algebra methods. This naturally leads to the RHSs transforming into rather cumbersome expressions so we omit displaying them here.

With the boundary conditions for the inner and outer homogeneous solutions computed, we numerically solve the coupled set of three ODEs as before. The vector space of linearly independent homogeneous solutions is now 6-dimensional and is constructed from inner, outer homogeneous solutions generated using BC obtained from the sets $\{b^3_0, b^3_1, b^5_1 \} $ for the inner and $\{a^3_\el, a^5_\el, a^5_{\el+2} \} $ for the outer solutions, respectively. So at each $(\el \ge 2 ,m=0) $ even mode, we numerically integrate the ODEs for a total of $2\times 6= 12 $ times. To determine the inhomogeneous solutions $ R^{(1),(3),(5)}(r_0) $ and their inner/outer $r$-derivatives, we construct a $6\times 6 $ complex matrix and invert it to solve for the complex amplitudes  $x_1, \ldots, x_6 $ as before. We omit the details here as we have illustrated how to do this for both the generic odd and even modes in sections \ref{sec:Inhomog}, \ref{sec:even_inhomo} respectively. Once these fields are known, we can then use the gauge equations to construct $ R^{(6)}(r_0) $ and $ R^{(7)}(r_0) $ and their inner/outer $r$-derivatives at $r=r_0$.


\section{Computing The Gravitational Self-Force} \label{sec:GSF}
With all the metric fields $\b{h}^{(i)}$ and their $t,r$-
derivatives computed, we now focus on the actual calculation of
the gravitational self-force. We follow the prescription of
\cite{BSII} and \cite{BSI}.

Because we are modeling the small mass $\mu$ as a point particle,
we are faced with the issue of the divergence of the GSF at the
location of the particle. This requires a careful regularization
of the GSF to remove the divergent, but non-physical, piece from
it. We can write the regularized GSF as \cite{BaMiNaOSa}
\be F^\alpha (x_0) = \lim_{x\rightarrow x_0} \left[
F^\alpha_{\mrm{full}} (x) - F^\alpha_{\mrm{dir}} (x) \right],
\label{eq:F_reg} \ee
where $ F^\alpha_{\mrm{full}} $ is the ``full'' GSF constructed
from the metric perturbation, and $ F^\alpha_{\mrm{dir}} $ is the
``direct'' (divergent) piece of it. Physically speaking, $
F^\alpha_{\mrm{dir}} $ can be thought of as representing the
instantaneous part of the GSF that propagates along the past
light-cone of the particle.

In the mode-sum scheme, $ F^\alpha_{\mrm{full}} $ and $
F^\alpha_{\mrm{dir}} $ are decomposed into multipole modes $
F^{\alpha\: l}_{\mrm{full}} $ and $ F^{\alpha\:l}_{\mrm{dir}} $.
Thanks to this multipole expansion, the individual $l$-modes of
the divergent piece $F^\alpha_{\mrm{dir}} $ all have finite values
at the $x^\mu \rightarrow x^\mu_0 $ limit. $l$ here represents the
scalar spherical harmonic modes and it should not be confused with
the tensorial modal index $\el$ of the previous
sections.

Individual $l$-modes of $ F^\alpha_{\mrm{full}} $ are obtained
from the fields $ \bar{h}^{(i)\el m}$ and their derivatives as
given by Eq.(\ref{eq:F_full2}) below. Then the GSF at the location
of the particle ($x_0$) is given by
\be F^\alpha (x_0) = \sum_{l=0}^\infty\: \left(
[F^{\alpha\:l}_{\mrm{full}}(x_0)]_\pm -A^\alpha_\pm L_{1/2} -
B^\alpha \right) \ \equiv \sum_{l=0}^\infty [F^{\alpha\:
l}_{\mrm{reg}}(x_0)]_\pm \ , \label{eq:F_reg_l_mode_sum} \ee
where $ L_{1/2} \equiv l +1/2 $. The $\pm $ correspond to taking
the $r$-derivative at the $ r\rightarrow r_0^\pm $ limit. $
A^\alpha_\pm $ and $ B^\alpha $ are regularization parameters.
They are derived from the local structure of $
F^{\alpha\:l}_{\mrm{dir}} $ near $x^{\mu\: \pm}_0$. $  \mp
A^\alpha_\pm L_{1/2} + B^\alpha$ represents the asymptotic form of
$ F^{\alpha\:l}_{\mrm{dir}} $ for large $l$. For circular orbits
in Schwarzschild spacetime, $ A^\alpha_\pm = B^\alpha = 0 $ for
$\alpha = t,\theta, \phi $. The non-zero $r$-components are given
by \cite{BaMiNaOSa}
\ba
A^r_\pm & = & \mp \f{\mu^2}{r_0^2} \left(1-\f{3M}{r_0} \right)^{1/2} \label{eq:A_mu}, \\
B^r & = & \f{\mu^2 r_0 \tilde{E}_0^2}{\pi ( \tilde{L}_0^2 + r_0^2)^{3/2} } \left[\hat{E}(w)-2\hat{K}(w) \right], \label{eq:B_mu}
\ea
where $\tilde{L}_0 = (Mr_0)^{1/2}/(1-3M/r_0)^{1/2}$ is the orbital angular momentum, $ \hat{K}(w) \equiv \int_0^{\pi/2} (1-w \sin^2 x )^{-1/2} dx $ and $ \hat{E}(w) \equiv \int_0^{\pi/2} (1-w \sin^2 x )^{1/2} dx    $ are the complete elliptic integrals of first and second kind, respectively and $ w \equiv (r_0/M-2)^{-1} $. The regularized GSF can be computed by using either one of the $\pm$ values: the quantity $ F^{\alpha l \:\pm}_{\mrm{full}} - L_{1/2} A^\alpha_\pm $ is direction independent. This $\pm$ equality provides us with a way to check our GSF results. Since the $t$-component needs no regularization, we can write $ F^{t \:l}_{\mrm{reg}\: \pm} = F^{t\: l}_{\mrm{full} \:+} = F^{t\:l}_{\mrm{full}\:-} $.

The $l$ modes of the full force are given by \cite{BSI}
\ba \left[F^{\alpha \: l}_{\mrm{full}}(x_0)\right]_\pm & = & \f{\mu^2}{r_0^2}\sum_{m=-l}^l Y^{l m}\left(\pi/2, \phi_0\right)\times \label{eq:F_full2}  \\
& \: & \left[ \mathcal{F}^{\alpha\: l-3, m}_{(-3)} + \mathcal{F}^{\alpha\: l-2, m}_{(-2)} + \mathcal{F}^{\alpha\: l-1, m}_{(-1)} + \mathcal{F}^{\alpha\: l, m}_{(0)} + \mathcal{F}^{\alpha\: l+1, m}_{(+1)} +\mathcal{F}^{\alpha\: l+2, m}_{(+2)} +\mathcal{F}^{\alpha\: l+3, m}_{(+3)} \right] . \nn
\ea
$\mathcal{F}^{\alpha l m}_{(j)} $ are constructed from $ \b{h}^{(i) \el m},\:  \b{h}^{(i) \el m}_{,r \pm},\:  \b{h}^{(i) \el m}_{,t} $ at $ x^\mu = x^\mu_0 $.  The expressions for $ \mathcal{F}^{\alpha l m}_{(j)}$ are quite lengthy and are explicitly given in appendix C of \cite{BSI} for circular and in appendix C of \cite{BSII} for eccentric orbits in Schwarzschild geometry. For this reason, we omit presenting them here. However, we would like to remark that $\mathcal{F}^{\alpha l m}_{(j)} $ contain coupling terms between tensor modes $\el$ and scalar modes $l$. This is because the metric perturbation $\bar{h}_{\mu\nu}$ is decomposed in terms of \emph{tensor} modes $\el$, but the GSF is computed by summing over \emph{scalar} modes $l$ (the regularization procedure requires the mode decomposition to be done in spherical harmonics \cite{BOI}, \cite{BOII}). As a result, a given scalar spherical harmonic mode $ l$ will couple to 5 tensor spherical harmonic modes with $ \el-2 \le l \le \el+2 $ for the $r$-component, and to 7 tensor modes $ \el-3 \le l \le \el+3 $ for the $t$-component of $ \mathcal{F}^{\alpha\: l}_{(j)}$. This is the reason why the index $(j)$ in Eq.(\ref{eq:F_full2}) goes from $(-3)$ to $(3)$.

An extra simplification arises in Eq.(\ref{eq:F_full2}) because the spherical harmonics $ Y^{l m}(\pi/2, \phi_0) = 0 $ for $l-m = \mrm{odd} $. Furthermore, because $\bar{h}^{(i)} Y^{l m} \rightarrow [\bar{h}^{(i)} Y^{l m}]^\ast $ under $ m \rightarrow -m $, we compute the sum only from $ m=1 $ to $ m=\el $ then fold over the $m$-sum properly to include the $ m < 0 $ contribution and finally add to these the $m=0 $ term in the summation in Eq.(\ref{eq:F_full2}). This is then regularized at each $l$ mode via Eq. (\ref{eq:F_reg_l_mode_sum}).

To obtain the final value for the GSF, we compute the sum over all
scalar $l$ modes.  Since the $t$-component converges
exponentially, $l_{\mrm{max}} \approx 10 $ suffices to obtain the
value of $F^t(x_0)$ to machine accuracy. However, the
$r$-component of the GSF falls off as $L_{1/2}^{-2} $ and this
converges much more slowly. As we are using finite computer power
to calculate an infinite sum over $l$, we must truncate the sum
for the $r$-component at some $l=l_{\mrm{max}} $ (usually
somewhere between $15$ and $30$) and use fitting methods to
estimate contribution from the $l > l_{\mrm{max}}$ modes. This
contribution accounts for at most $\sim 2\% $ to the overall GSF
\cite{BSI} and is called ``the large-$l$ tail''. The details of
how to compute it are given extensively in section IIIE of
\cite{BSI}. Basically, one extrapolates the $l>l_{\mrm{max}}$
terms in the sum using polynomial fits in powers of
$L_{1/2}^{-2}$. As we use the same fitting method as \cite{BSI},
we refrain from elaborating any further. The details can be found
there but let us discuss briefly how the tail error depends on the
parameters used to do the fit.

There are two free parameters that determine the large-$l$ tail.
The first one is the number $k$ of $l$ modes $ \in
[l_{\mrm{max}}+1-k, l_{\mrm{max}}] $ that we select for the
extrapolation. The second is $ N$, which determines the degree of
the polynomial fit in powers of $ L_{1/2}^{-2} $. We use a numerical scheme that
varies these two parameters ($k,N$) and finds the optimal values
for both by comparing the error between the regularized $l$ modes
$ F^{r\:l_{\mrm{max}}+1-k \le l  \le l_{\mrm{max}}}_{\mrm{reg}} $
obtained from the fitting formula and the actual numerical values
computed by solving the Einstein equations. Our scheme uses the
following ranges for the two parameters: $ 2 \le N \le 6 $ and $ 5
\le k \le 12$ depending on the total number of $l$ modes that we
compute (varies from 15 to 30). Because our frequency-domain code
is able to compute up to $30$ modes within an hour for $r_0 < 20M
$, we are able to reduce the fractional error in the tail
computation to $\sim 10^{-8} $. As we will see below, the
uncertainty in the large-$l$ tail is not always the source of the
most significant error in our computation.

\subsection{Summary of Methods and Computational Details}
Working in the frequency domain, we started by numerically solving
the 10 coupled field equations (\ref{eq:R_field_eq}) for the
radial fields $ R^{(i)}_{\el m}(r) $ (the modes $(\el,m) = (0,0),
(\mrm{odd},0)$ have analytic solutions). To this end, for the
first time, we calculated the boundary conditions for the radial
fields in Lorenz gauge. We constructed linearly independent bases
of homogeneous solutions and used these to obtain the
inhomogeneous solutions $R^{(i)}_{\el m}(r_0)$ and their
$r\rightarrow r_0^\pm $ $r$-derivatives via junction conditions.
Following the prescription of \cite{BSI}, we computed the
$\mathcal{F}^{t\:lm \pm}_{(j)}, \mathcal{F}^{r\:lm \pm}_{(j)}$ of Eq.
(\ref{eq:F_full2}). The $l$ modes  of the `full' GSF are then
given by this equation. We regularized the GSF at each $l$ mode
with the help of Eq. (\ref{eq:F_reg_l_mode_sum}) then added all
the individual $l$-mode contributions together. Finally, for the
$r$-component, we added the large-$l$ tail to the $l$ sum to
account for the $F^{r\:l > l_{\mrm{max}}}_{\mrm{reg}}$ terms that we did not
actually compute. It is this final result that equals the actual
gravitational self-force. It is this quantity that we compare with
BS in section \ref{sec:results}.

Our numerical code is written in C and uses Gnu Scientific Library
(GSL) repositories \cite{GSL} for the numerical integration of the
ODEs and matrix algebra used in obtaining the inhomogeneous
solutions. After exhaustive numerical experimentation, we selected
to work with the Runge-Kutta Prince-Dormand (rk8pd) numerical
integration routine as this proved to be the fastest. For our
matrix inversion, we opted for the lower-upper (LU) triangular
matrix decomposition. We use a single desktop machine with two
quad-cores to run our code, which proved to be more than
sufficient for GSF computations for circular orbits. More than
95\% of the computing time is taken up by the numerical
integration of the coupled ODEs. This task is further multiplied
because of the need to construct $N$-dimensional bases of
homogeneous solutions. For example, a GSF computation due to the
first 15 scalar modes (i.e. tensor $ \el = 0 \ldots 18 $)
numerically integrates various coupled ODEs a total of $2192$
times.

The speed of the numerical ODE integrator depends on a few freely specifiable parameters: the size of the integration domain $[r^\ast_{in}, r^\ast_{out}] $,   and the numerical accuracy thresholds ($\Delta_{\mrm{rel}}, \Delta_{\mrm{abs}}$) used by the integrator. Given an ODE, the code picks the smaller of the two thresholds to integrate. We have empirically determined that a relative ODE solver accuracy of $\Delta_{\mrm{rel}} =10^{-10}$ is sufficient for computing the GSF to within an overall fractional error of $\lesssim 10^{-6}$ for runs with orbital radii $6M \le r_0 \le 50M$. However, for $r_0 > 50M$ runs, we observed that $\Delta_{\mrm{rel}}$ needs to be brought as close to machine accuracy as reliably possible i.e. $10^{-14}$.  This is because the transition region between the outer wave-zone (where the homogeneous fields $\bar{h}^{(i)} \rightarrow \mrm{e}^{-i \omega_m (t-r_\ast)} $) and the region where the fields exhibit power-law growth (near $r_0$) is farther out for larger $r_0$. Therefore, the numerical solutions can possibly grow by more than 20 orders of magnitude as the routine integrates from $r_{\mrm{out}}$ to $r_0$. This fundamentally limits the accuracy that we can reach with a numerical integrator using double floating point precision. After some numerical experimentation, we settled on a scheme that adaptively varies $\Delta_{\mrm{rel}}, \Delta_{\mrm{abs}}$ with increasing $r_0$. The scheme works well for up to $r_0 =100M $ beyond which the accuracy thresholds thread very close to machine accuracy and the runtimes grow unreasonably long.

The runtimes are rather insensitive to the location of $ r^\ast_{\mrm{in}}$. The reason is that the potential $ V_{\el m}$ is very `flat' near the event horizon (less than $1\%$ variance as one goes from $r_{in}^\ast=-35$ to $-55 $), so the solutions hardly change. On the other hand, the runtimes do depend heavily on the location of $r^\ast_{\mrm{out}} $. Therefore, its location must be chosen carefully. We elaborate more on this in the next subsection.

\subsection{The Error Budget}\label{sec:errors}
The major sources of error that go into our computation are: (1) Error in the large-$l$ tail, (2) Error in the numerical matrix inversions used to construct the inhomogeneous solutions, (3) Numerical discretization error in the numerical integration of the ODEs, and (4) The fact that the boundary conditions are not computed at $r^\ast = \pm \infty $.

We determined that the error coming from the finiteness of the
locations of the boundary points is much smaller than the other
three sources of error. After some numerical experimentation, we
came up with a satisfactory location for $r_{\mrm{out}} $
($r^\ast_{\mrm{out}}$) keeping in mind the wave-zone condition
$r_{\mrm{out}}
>> \el r_0/\omega_m $ and the fact that our code slows down too
much if $r_{\mrm{out}}$ is unnecessarily too far out. This optimal
choice was mentioned earlier in section \ref{sec:odd_sec}. We
tested the sensitivity of our solutions against changing
$r_{\mrm{out}}$. We found that the relative variation in
$|\b{h}^{(i)}|$ was $\lesssim \mathcal{O}(10^{-12}) $ when
$r_{\mrm{out}} $ was increased by up to one order of magnitude.

We have already commented on the errors in the large-$l$ tail computation. Our usual standard has been a fractional error of $10^{-6}$ in the large-$l$ tail. As mentioned in section \ref{sec:GSF}, we can reduce this error to nearly $1.0 \times 10^{-8}$ by computing more numerical modes, but this naturally increases the runtimes. On the other hand, if we adhere to a fractional error of $10^{-4}$ or $ 10^{-5}$ then we can reduce the overall runtimes considerably by computing less modes. We show this in Fig. \ref{fig:runtimes}, where we display plots of runtimes vs. $r_0$ for overall fractional errors of $ 10^{-4}, 10^{-6}$ and $10^{-7}$. In short, we have a good understanding and good control over the uncertainty in the large-$l$ tail.

The numerical discretization error coming from the numerical
integration of the ODEs contributes much less to the overall error
than the other error sources mentioned here. The GSL ODE
integrator routines are very robust and have a very good handle on
discretization errors. Our own numerical tests showed that these
errors have magnitudes $ \lesssim \mathcal{O}(10^{-12}) $ with
respect to the inhomogeneous fields.

Finally, as mentioned in section \ref{sec:even_inhomo}, the
biggest source of error comes from the numerical inversion of the
matrix constructed from the homogeneous solutions. This becomes
the dominant source of error for $r_0 \gtrsim 50M $. An inspection
of the matrix inversion output for each ($\el,m$) mode reveals
that the inversion errors grow with increasing $ \el-m $ and that
they are also larger in the even parity sector. We monitored the
condition numbers of the matrices and found out that for even
parity modes with $\el-m > 15 $, they routinely exceeded $10^{12}$
for $r_0>50M$ and got as large as $10^{22} $ for $ r_0 >100M $.
Further inspection of these large $\el$, large $r_0$ even modes
revealed that the determinant threads very close to zero. This is
an indication that our linearly independent bases of homogeneous
solutions start becoming degenerate in this region. The reason why
this happens for large $\el-m$ is due to particular way we have
formulated the location of the outer boundary by setting
$r_{\mrm{out}} = 50\: \el r_0/\omega_m = 50 \: r_0^{5/2} (\el/m)
$. From this, one sees that $r_{\mrm{out}}$ reaches its maximum
value when $\el-m$ reaches its maximum value. So, this `degeneracy
problem' is actually caused by large values for $r_{\mrm{out}}$.
What happens is that because the leading order power-law for each
homogeneous field dominates near $r_0$, the solutions that have
the same power-law behavior start looking numerically identical as
the integrator works its way in toward $r_0$. As we look at the
values of the fields for larger $r_0$ runs, the matrices
constructed from the even parity homogeneous fields become
linearly dependent (singular valued). This means the matrix
inversion is not very reliable.
We find that this degeneracy of even parity solutions becomes
significant for the runs where $r_0 \gtrsim 50M$. So, any
numerical ODE integration that routinely goes beyond this point
($r_0 \approx 50M$) starts running into this degeneracy problem.

We model the error coming from the singular-valuedness of the
matrices as a continuity violation in the inhomogeneous fields
$R^{(i)}(r)$ at $r=r_0$ . This continuity violation,
$\Delta^{(i)}$, is most prominent for the fields $R^{(2),(4),(5)}$
where it is about $\mathcal{O}(10^4)$ larger than the
violations in the other fields. In the worst case, e.g. $r_0=150M$
and $\el =17,\: m=1$; $\Delta^{(5)} \approx 10^{-5} $. However,
even at $r_0=150M$, the violation quickly subsides to $\lesssim
10^{-9}$ once $ m \ge 2 $ whatever $\el $ may be, but because the
GSF is constructed by summing over all ($\el,m$) modes, this error
is additive. For a computation of the GSF requiring
$\el_{\mrm{max}} = 18$, the relative strength of the error is
amplified by a factor of $\sim 10^2 - 10^3 $ going from
from a single mode to the final GSF, which is constructed from the sum of $\mathcal{O}(10^2)$ modes. This is
indeed what we observe numerically. We have not yet looked into
fixing this inversion problem but we are aware that using
singular-valued decompositions for the matrices do not offer an
improvement \cite{Bar_PVT}. Be that as it may, we do not think
this to be a problem for when we compute the GSF for eccentric
orbits because we will be mostly interested in the strong field
regime of $r_0 < 20M $. However, for equatorial eccentric orbits
in frequency domain, we expect to encounter a similar type of
degeneracy in our solutions due to the fact that the frequency
spectrum is determined by two fundamental frequencies:
$\omega_{mn} = m \Omega_\phi + n \Omega_r $. There will be points
in the parameter space where the two terms in $\omega_{mn}$ will
conspire to cancel each other to values less than $10^{-4}$. When
this happens, the conditions numbers for matrices of homogeneous
solutions grow to values that render the matrix inversion
unreliable. We are currently working on a solution to this
problem.

\section{Results}\label{sec:results}
We present the output of our frequency-domain code for the
gravitational self-force in Tables 2 and 3. For comparison, we
include the results of BS \cite{BSI} and the relative difference
between our respective values for the $t$-, and $r$-components of
the GSF. We find very good agreement with the results of BS
(within their error bars) for $r_0$ up to $\sim 100M$. However,
beyond that, our values stray from theirs. Given that Berndtson
\cite{Berndtson2009} agrees with BS within their quoted errors
bars for up to $150M$, we must conclude that the degeneracy
problem renders our results unreliable beyond $r_0 \sim 100M$.
However, as our results show, in the strong field regime our
f-domain results are much more accurate than their time-domain
counterparts.

As another way of confirming our results and determining the
magnitude of the error in our GSF computation, we computed the
energy flux of the gravitational waves leaving the system and
compared the value of the total radiated power with the total rate
of energy loss given by the dissipative component of the GSF. In
the case of circular orbits, only the $t$-component of the GSF is
dissipative so the rate of energy loss can be related to $F^t$ as
follows:
\be \f{d\tilde{E_0}}{d\tau} = -\mu^{-1} F_t .
\label{eq:energy_flux1} \ee
In terms of Schwarzschild time $t$, this
becomes $ d\tilde{E_0}/dt = - (\mu u^t_0)^{-1} F_t $, where
$u^t_0$ is the $t$-component of the 4-velocity of the particle
evaluated at $r=r_0$. In the adiabatic approximation, where $\mu/M
\ll 1 $, $d\tilde{E_0}/dt$ can be taken to be the average rate of
energy loss per orbit. Energy conservation dictates that this loss
of energy must be balanced by the total energy flux carried by
gravitational waves radiated out to infinity and absorbed into the
black hole. Therefore, we have the following balance equation: \be
\dot{E}_{\mrm{total}} \equiv \dot{E}_\infty + \dot{E}_{\mrm{EH}} =
-\mu \f{d\tilde{E_0}}{dt} = F_t/ u^t_0, \label{eq:energy_balance}
\ee where the overdot now denotes $d/dt$ and $ \dot{E}_\infty,
\dot{E}_{\mrm{EH}} $ denote the gravitational wave flux radiated
to infinity and through the event horizon (EH), respectively.
These fluxes are constructed from the metric fields $ \bar{h}^{(i)
\el m} $. We omit the details of this construction here, but for
the interested reader they can be found in \cite{BSI, Teu1, Teu2}.
Let us simply display the final expressions for the fluxes:
\be
\dot{E}_\infty = \sum_{\el=2}^\infty \sum_{m=-\el}^\el \f{\mu^2 m^2 \Omega_0^2}{64 \pi \lambda \el (\el+1)} \left| \bar{h}^{(7)}_\infty - i \bar{h}^{(10)}_\infty \right|^2, \label{eq:E_flux_inf}
\ee
\ba
\dot{E}_{\mrm{EH}} & = &\sum_{\el=2}^\infty \sum_{m=-\el}^\el \f{\mu^2 \lambda \el (\el+1)}{256 \pi M^2 (1+16 M^2 m^2 \Omega_0^2)} \label{eq:E_flux_EH} \\
\quad & \quad & \times \left| \bar{h}^{(1)}_{\mrm{EH}} + \f{1+4iMm \Omega_0}{\el(\el+1)} \left[ \bar{h}^{(5)}_{\mrm{EH}} - i \bar{h}^{(9)}_{\mrm{EH}} +2i M m \Omega_0 \lambda^{-1} \left( \bar{h}^{(7)}_{\mrm{EH}} - i \bar{h}^{(10)}_{\mrm{EH}} \right) \right]\right|^2 \nn ,
\ea
where $\bar{h}^{(i)}_{\infty, \mrm{EH}} $ implies that the metric field is to be evaluated  (in frequency domain) at $ r=\infty, 2M $, respectively. Using Eqs. (\ref{eq:E_flux_inf}), (\ref{eq:E_flux_EH}) and our results for $F^t$, we compute the total radiated power $\dot{E}_{\mrm{total}}$ and compare the resulting values. The relative difference between the two results is shown in the last column of Table 3, which shows that the agreement is excellent for small $r_0$. It naturally gets worse for increasing values of $r_0$. We also found that the disagreement between the two values for $\dot{E}_{\mrm{total}}$ matched our overall fractional error in $F^t$ well.

\begin{figure}
     \centering
     \subfigure{
    \label{fig:1e4_runtimes}
          \includegraphics[width=.45\textwidth]{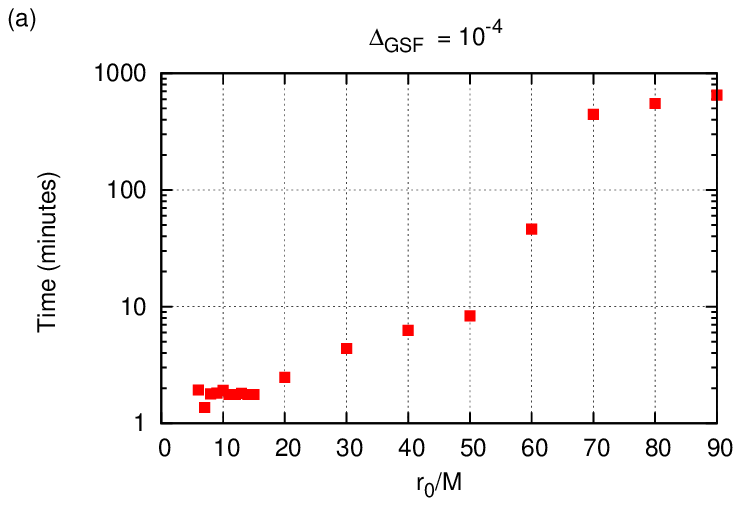}}
     \hspace{.3in}
    \subfigure{
          \label{fig:1e6_runtimes}
          \includegraphics[width=.45\textwidth]{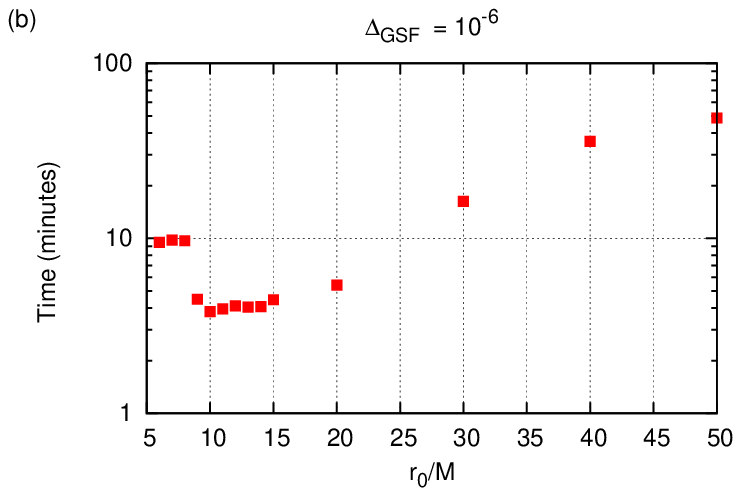}}\\
     \subfigure{
    \label{fig:1e7_runtimes}
          \includegraphics[width=.45\textwidth]{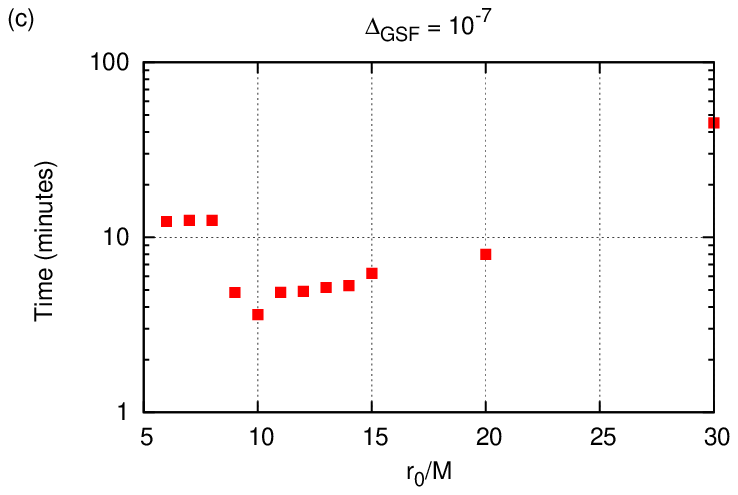}}
     \hspace{.3in}
    \subfigure{
          \label{fig:strong_field_runtimes}
          \includegraphics[width=.45\textwidth]{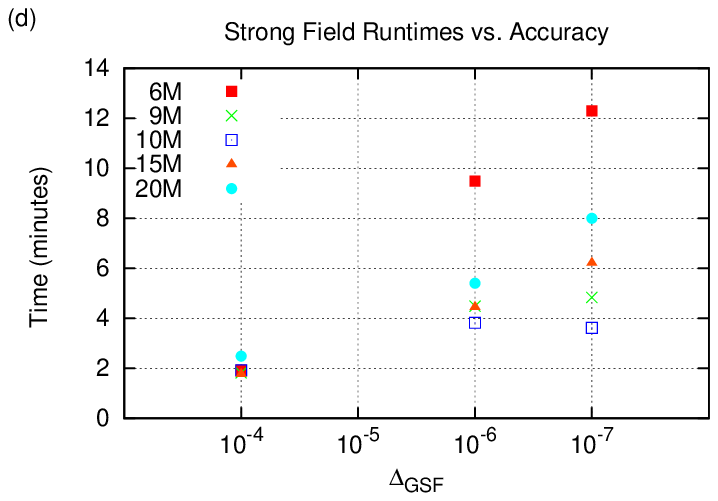}}
    \caption{The runtimes for the $ 10^{-4}, 10^{-6}, 10^{-7}$ overall fractional error runs. Panels (a), (b) and (c) display plots of runtime (in minutes) versus orbital radius $r_0$ at which we compute the GSF. $\Delta_{\mrm{GSF}}$ denotes the overall fractional error in our numerical computation of the GSF. This error is what we refer to as our (relative) `accuracy'. As can be seen in panel (a), at an accuracy of $10^{-4}$, our code takes less than two minutes to compute the GSF for  $r_0 \lesssim 15M$. This grows nearly to a day as $r_0$ approaches $100M$. Panel (b) shows that an accuracy of $10^{-6}$ increases the runtimes by a factor of two to three for $r_0 \lesssim 10M $, but the runtimes are still $\lesssim 10$ minutes for $r_0 \lesssim 20M$. However, beyond $r_0 = 50M$, this accuracy becomes unattainable. As panel (c) shows, an accuracy of $10^{-7} $ is achievable for $r_0 \lesssim 30M $ and the overall runtimes do not change much for these strong field GSF computations. Interestingly enough, for $r_0 \lesssim 20M$, the $r_0 \le 8M $ runs seem to take more time than $r_0 \ge 9M$ runs. This is a result of our having to compute more modes to obtain the GSF for the $r_0 \le 8M $ runs because the large-$l$ tail could not be computed to the desired accuracy of $10^{-6}$ or $10^{-7}$ using just 17 scalar modes, which is what we had done for the $r_0 \ge 9M$ runs.
We think the reason for this is that the magnitudes of the
individual $l$ modes of the GSF are large enough for $r_0 \le 8M $
that more modes are needed in order for the tail to be fit
correctly. Finally in panel (d), we present the runtimes for a few
$r_0 \le 20M $ run for all three accuracies. As expected, the
runtimes increase with demand for higher accuracy (except for the
$10M$ run). Most importantly, the figure shows that all $r_0 \le
20M$ runs take less than 15 minutes up to an accuracy of
$10^{-7}$.}
     \label{fig:runtimes}
\end{figure}

We also present the runtimes for our code for three different
relative accuracies. These are quantified by the overall
fractional error in our numerical computation of the GSF. We have
selected to present results for overall fractional errors of $
10^{-4}, 10^{-6}, 10^{-7}$. We display the runtimes for these in
Fig. \ref{fig:runtimes}. As can be seen from the upper left panel
of the figure, at a relative accuracy of $10^{-4}$, our code takes
less than two minutes to compute the GSF for radii less than $\sim
15M$. This grows nearly to a day as $r_0$ approaches $100M$.
Although toward $100M$ the runtimes appear to level off, this is
due to our logarithmic scale for the vertical axis. The runtimes
increase by $\sim 100$ minutes in going from $70M$ to $80M$, and
$80M$ to $90M$. In the same figure, upper right panel, one sees
that demanding an accuracy of $10^{-6}$ increases the runtimes by
a factor of two to three for $r_0 \lesssim 10M $. However, beyond
$r_0 = 50M$, this accuracy becomes unattainable. Finally, we find
it quite difficult to keep the overall fractional error less than
$10^{-7} $. But as the lower left panel of the figure shows, an
accuracy standard of $10^{-7}$ is achievable for $r_0 \lesssim 30M
$ and the overall runtimes are not prolonged by much for these
strong field GSF computations. Interestingly enough, in the regime
$r_0 \lesssim 20M$, the $r_0 \le 8M $ runs seem to take more time
than $r_0 \ge 9M$ runs. This was artificially caused by our need
to compute more modes in order to lower the error in the large-$l$
tail for the $r_0 \le 8M $ runs. It turns out that for the
smallest radii, the large-$l$ tail can not be computed to the
desired accuracy of $10^{-6}$ or $10^{-7}$ using just 15 or 17
scalar modes, which is what we had done for the $r_0 \ge 9M$ runs.
We think the reason for this is that the magnitudes of the
individual $l$ modes of the GSF are large enough for $r_0 \le 8M $
that more modes are needed in order for the tail to be fit
correctly. Finally, in the lower right panel, we present the
computation times for a given $r_0 \le 20M $ run for all three
accuracies. As expected, the runtimes increase with demand for
higher accuracy. However, by how much they increase is not the
same at each radius. There is also the anomalous data point for
the $10M$ run where the $10^{-7}$ accuracy computation takes
slightly less time than the $10^{-6}$ one. This comes from our not
having explored thoroughly enough the free parameters that
determine the overall error and runtime such as $l_{\mrm{max}}$,
number of points used in the tail and the numerical ODE integrator
accuracy thresholds.
Most importantly, the figure shows that all $r_0 \le 20M$ runs
take less than 15 minutes up to an accuracy of $10^{-7}$.

It should also be added that even on our modest desktop, we can
simultaneously perform a dozen strong field runs without
significantly affecting individual runtimes. For example, in a 15
minute period, we can compute the GSF for all integer orbital
radii from $6M$ to $10M$ to an accuracy of $10^{-6}$. We find the
speed of our code to be fast enough to encourage continuing this
frequency-domain approach to tackle the eccentric Schwarzschild
problem for the GSF. Work is currently underway and the
preliminary results are encouraging. We intend to apply these
methods to the full Kerr problem later on.

\begin{tiny}
\begin{table}
\begin{tabular}{|c|c|c|c|c|c|c}  \hline \hline
        $ r_0/M $ & $ (M/\mu)^2 F^r  $ & $ (M/\mu)^2 F^r_{BS} $ & Rel. diff.  \\ \hline
   6.0 & $ 2.4466495(4)\times 10^{-2} $  & $2.44661\times 10^{-2} $  & $ 4.0 \times 10^{-6} $\\
    7.0 & $2.149907776(8)\times 10^{-2} $  & $2.14989\times 10^{-2} $  & $8.3 \times 10^{-6} $ \\
    8.0 & $1.8357830(4)\times 10^{-2} $  & $1.83577\times 10^{-2} $  & $7.1\times 10^{-6} $ \\
    9.0 & $1.5637099(1)\times 10^{-2} $  & $1.56369\times 10^{-2} $  & $1.3\times 10^{-5} $ \\
    10.0 & $ 1.3389470(2)\times 10^{-2} $  & $1.33895\times 10^{-2} $  & $2.2 \times 10^{-6} $  \\
    11.0 & $1.155174593(6)\times 10^{-2} $  & $1.15518\times 10^{-2} $  & $4.7 \times 10^{-6} $  \\
    12.0 & $1.00462381(8)\times 10^{-2} $  & $1.00463\times 10^{-2} $  & $ 6.2 \times 10^{-6} $ \\
    13.0 & $8.8048853(3)\times 10^{-3} $  & $ 8.80489\times 10^{-3} $ & $ 5.3 \times 10^{-7} $ \\
    14.0 & $7.7730602(4)\times 10^{-3} $ & $7.77307\times 10^{-3} $ & $1.3\times 10^{-6} $ \\
    15.0 & $6.9081719(3)\times 10^{-3} $ & $6.90815\times 10^{-3} $ & $9.7 \times 10^{-5} $\\
    20.0 & $4.1570550(2)\times 10^{-3} $ & $4.15706\times 10^{-3} $ & $ 1.2\times 10^{-6} $  \\
    30.0 & $1.9698169(3)\times 10^{-3} $ & $1.96982\times 10^{-3} $ & $1.6 \times 10^{-6} $  \\
    40.0 & $1.142883(1)\times 10^{-3} $ & $1.14288\times 10^{-3} $ & $ 2.6\times 10^{-6} $   \\
    50.0 & $7.449480(1)\times 10^{-4} $ & $7.44949\times 10^{-4} $ & $1.3\times 10^{-6} $  \\
    60.0 & $5.236083(3)\times 10^{-4} $ & $5.23613\times 10^{-4} $ & $9.0\times 10^{-6} $ \\
    70.0 & $3.8801(1)\times 10^{-4} $ & $3.88010\times 10^{-4} $ & $2.6\times 10^{-6} $ \\
    80.0 & $2.9896(1)\times 10^{-4} $ & $2.98979\times 10^{-4} $ & $6.4\times 10^{-5} $  \\
    90.0 & $2.3739(1)\times 10^{-4} $ & $2.37406\times 10^{-4} $ & $6.7 \times 10^{-5} $ \\
    100.0 & $1.9304(1)\times 10^{-4} $ & $1.93063\times 10^{-4} $ & $1.2 \times 10^{-4} $ \\
    120.0 & $1.3483(1)\times 10^{-4} $ & $1.34868\times 10^{-4} $ & $2.8 \times 10^{-4} $ \\
    150.0 & $8.673(1)\times 10^{-5} $ & $8.68274\times 10^{-5} $ & $1.1 \times 10^{-3} $
 \\ \hline
    \end{tabular}
\caption{Output for the $r$-component of the gravitational
self-force for various orbital radii $r_0$ compared with results
of BS \cite{BSI}. Column 2 contains our results; the number in
parentheses indicates the size of the uncertainty in the last
significant digit, e.g. $2.4466495(4) = 2.4466495 \pm 4\times
10^{-7}$. In column 3, we display the results of BS for
comparison. Column 4 gives the relative difference between our
values and BS'. Our results are within their quoted error bars for
nearly up to $r_0=100M$. Beyond that the disagreement seems to
grow up $\mathcal{O}(10^{-3})$. Given that Berndtson's results
\cite{Berndtson2009} agree with BS better for large $r_0$, we
conclude that our current results are not reliable beyond $r_0
\approx 100M$.
Nevertheless, as can be seen from the number of significant digits
that we have included for $F^r$ for $r_0 \lesssim 50M$, the
frequency-domain results are much more accurate than time domain
in the strong field regime.}
\end{table}
\end{tiny}

\begin{tiny}
\begin{table}
\begin{tabular}{|c|c|c|c|c|c|c}  \hline \hline
        $ r_0/M $ & $(M/\mu)^2 F^t  $ & $(M/\mu)^2 F^t_{BS}$ & Rel. diff. & $ (M/\mu)^2\dot{E}_{total} $ & $(M/\mu)^2 F_t/u^t_0 $ & Rel. diff.  \\ \hline
     6.0 & $ -1.9947610064(3) \times 10^{-3}$ & $-1.99476 \times 10^{-3} $& $ 5.0 \times 10^{-7} $ & $ 9.4033935631 \times 10^{-4}$ & $ 9.4033935626 \times 10^{-4}$ & $ 5.7 \times 10^{-10} $ \\
    7.0 & $-7.411127850(9)\times 10^{-4} $ & $ -7.41101\times 10^{-4}$ & $1.2 \times 10^{-5}$ & $ 4.001632906\times 10^{-4} $ & $ 4.001632909\times 10^{-4}$ & $ 6.6 \times 10^{-11} $ \\
8.0 & $-3.307397510(3)\times 10^{-4} $ & $ -3.30740\times 10^{-4}$ & $1.2 \times 10^{-5}$ & $ 1.9610454858\times 10^{-4} $ & $ 1.9610454864\times 10^{-4}$ & $ 3.0 \times 10^{-10} $ \\
9.0 & $-1.668101230(4)\times 10^{-4} $ & $ -1.66810\times 10^{-4}$ & $1.2 \times 10^{-5}$ & $ 1.0593325177\times 10^{-4} $ & $ 1.0593325178\times 10^{-4}$ & $ 8.8 \times 10^{-11} $ \\
    10.0 & $ -9.19075772(7)\times 10^{-5}$ & $ -9.19067\times 10^{-5}$ & $ 1.0 \times 10^{-5} $ & $ 6.151631678\times 10^{-5} $ & $ 6.151631677 \times 10^{-5} $ & $ 2.2 \times 10^{-10} $ \\
11.0 & $-5.41623002(6)\times 10^{-5} $ & $ -5.41605\times 10^{-5}$ & $1.2 \times 10^{-5}$ & $ 3.779162580\times 10^{-5} $ & $ 3.771962578\times 10^{-5}$ & $ 4.8 \times 10^{-10} $ \\
12.0 & $-3.3659568(1)\times 10^{-5} $ & $ -3.36587\times 10^{-5}$ & $1.2 \times 10^{-5}$ & $ 2.42917009\times 10^{-5} $ & $ 2.42917010\times 10^{-5}$ & $ 3.2 \times 10^{-9} $ \\
13.0 & $-2.1839249(2)\times 10^{-5} $ & $ -2.18388\times 10^{-5}$ & $1.2 \times 10^{-5}$ & $ 1.620747493\times 10^{-5} $ & $ 1.620747489\times 10^{-5}$ & $ 2.8 \times 10^{-9} $ \\
14.0 & $-1.4685410(2)\times 10^{-5} $ & $ -1.46851\times 10^{-5}$ & $1.2 \times 10^{-5}$ & $ 1.115762106\times 10^{-5} $ & $ 1.115762104\times 10^{-5}$ & $ 2.2 \times 10^{-9} $ \\
15.0 & $-1.0177145(1)\times 10^{-5} $ & $ -1.01772\times 10^{-4}$ & $1.2 \times 10^{-5}$ & $ 7.88902019\times 10^{-6} $ & $ 7.88902015\times 10^{-6}$ & $ 5.3 \times 10^{-9} $ \\
    20.0 & $ -2.2554391(2)\times 10^{-6} $ & $ -2.25549 \times 10^{-6} $ & $ 2.2 \times 10^{-5} $ & $ 1.87147091\times 10^{-6}$ & $ 1.87147088\times 10^{-6} $ & $ 1.6 \times 10^{-8} $ \\
    30.0 & $ -2.8081894(8)\times 10^{-7} $ & $ -2.80813\times 10^{-7}$ & $ 2.1\times 10^{-5} $ & $ 2.486484\times 10^{-7} $ & $ 2.486486\times 10^{-7} $ & $ 1.1\times 10^{-6} $ \\
    40.0 & $ -6.51228(2)\times 10^{-8} $ & $ -6.51219\times 10^{-8} $ & $ 1.5\times 10^{-5} $ & $ 5.95014\times 10^{-8}$ & $ 5.95015\times 10^{-8}$ & $ 2.9\times 10^{-6} $ \\
    50.0 & $ -2.108456(4)\times 10^{-8} $ & $ -2.10849\times 10^{-8} $ & $ 3.0\times 10^{-5} $ & $ 1.962458\times 10^{-8} $ & $ 1.962453\times 10^{-8} $ & $ 2.3\times 10^{-6} $ \\
    60.0 & $ -8.41300(3)\times 10^{-9} $ & $ -8.41306\times 10^{-9} $ & $ 7.0\times 10^{-6} $ & $ 7.92644\times 10^{-9}$ & $ 7.92641\times 10^{-9} $ & $ 3.9\times 10^{-6} $ \\
    70.0 & $ -3.8743(1)\times 10^{-9} $ & $ -3.87411\times 10^{-9}$ & $ 4.1\times 10^{-5} $ & $ 3.6819\times 10^{-9} $ & $ 3.6818\times 10^{-9}$ & $ 3.5\times 10^{-5} $ \\
    80.0 & $ -1.9804(1)\times 10^{-9} $ & $ -1.98069\times 10^{-9} $ & $ 4.6\times 10^{-5} $ & $ 1.8945\times 10^{-9} $ & $ 1.8946\times 10^{-9} $ & $ 4.2\times 10^{-5}$ \\
    90.0 & $ -1.0966(3) \times 10^{-9} $ & $ -1.09654 \times 10^{-9} $ & $ 9.1\times 10^{-5} $ & $ 1.0541\times 10^{-9} $ & $ 1.0544 \times 10^{-9} $ & $ 2.6\times 10^{-4} $ \\
    100.0 &$ -6.464(2) \times 10^{-10}$ & $ -6.46305\times 10^{-9}$  & $2.1\times 10^{-4}$ & $ 6.238\times 10^{-10}$ & $ 6.240\times 10^{-10}$ & $ 3.2 \times 10^{-4} $ \\
        120.0 & $ -2.596(9)\times 10^{-10}$ & $ -2.59096\times 10^{-10}$ & $ 1.9\times 10^{-3} $ & $ 2.516\times 10^{-10}$ & $ 2.525\times 10^{-10}$ & $ 3.6\times 10^{-3} $  \\
    150.0 & $ -8.44(6)\times 10^{-11}$ & $-8.47172\times 10^{-11} $ & $ 3.4\times 10^{-3} $ & $ 8.27\times 10^{-11} $ & $ 8.22\times 10^{-11} $ & $ 6.1\times 10^{-3} $
 \\ \hline
    \end{tabular}
\caption{Output for the $t$-component of the gravitational self-force for
various radii $r_0$. Our results are in column 2. In column 3, we
show the results of BS for $F^t$ and display the relative
difference in column 4. Once again, our results fall within BS'
error bars for $r_0 \lesssim 100M$ and again the error increases
up to $\mathcal{O}(10^{-3})$ for $r_0=150M$. And as was the case
with $F^r$, our frequency-domain results for $F^t$ also have much
smaller uncertainties in the $r_0 \lesssim 50M$ regime compared
with those of BS.  We also checked our results for $F^t$ using
energy balance arguments. Since only the $t$-component of the GSF
is dissipative for circular orbits, it can be related to the
energy flux leaving the system as we have outlined in section
\ref{sec:results}. The total energy flux is computed using the two
different methods and these results are displayed in columns 5 and
6 down to the significant digit at which they start disagreeing.
Column 7 contains the relative difference between the two values.
Once again, the agreement is extremely good for small $r_0$ and
grows to $\mathcal{O}(10^{-3})$ as $r_0$ increases to $150M$.}
\end{table}
\end{tiny}

\section{Acknowledgements}
SA thanks Leor Barack, Nori Sago and Niels Warburton. This work was supported by STFC grant No. PP/E001025/1.

\appendix
\section{The Recursion Relations for The Boundary Conditions}\label{sec:recursion}
\subsection{Generic Odd and Even Modes}\label{sec:generic_rec}
Recall that by {\it generic}, we mean the non-static ($m\ne 0$), $\el > 1 $ modes. Here, $\omega$ denotes $\omega_m = m\Omega_0$ and $L \equiv \el(\el+1) $. We begin by redisplaying the recursion relations for the outer boundary conditions (BC) for odd parity homogeneous fields $R^\pm_9$ and $R^\pm_{10}$:
\ba
2 i \omega k\; a^9_k & =& C_{k-1}\: a^9_{k-1} + D_{k-2}\: a^9_{k-2} + E_{k-3}\: a^9_{k-3} + 2 a^{10}_{k-1} -10M a^{10}_{k-2} + 12M^2 a^{10}_{k-3} ,\nn \\ \label{eq:a9_k_A} \\
2 i \omega k\; a^{10}_k & = & I_{k-1}\: a^{10}_{k-1} + J_{k-2}\: a^{10}_{k-2} + K_{k-3}\: a^{10}_{k-3} + 2\lambda a^9_{k-1} - 4M\lambda a^9_{k-2} , \label{eq:a10_k_A}
\ea
where
\ba
C_k & = & 4M i\omega k + k(k+1) - L - 4, \qquad I_k = 4M i\omega k + k(k+1) - L + 2, \nn \\
D_k & = & -6Mk - 4M k^2 + 24M + 2ML, \quad J_k = -6Mk - 4M k^2 -6M + 2ML, \nn \\
E_k & = & 4 M^2 (k^2 + 2k-8), \qquad\qquad K_k = 4M^2 (k^2 +2k +1) . \nn
\ea
Next, we present the recursion relations for the inner BC
\ba
4M^2 k (k-4M i\omega) b^9_k & = & \tilde{C}_{k-1}\: b^9_{k-1} + \tilde{D}_{k-2}\: b^9_{k-2} + \tilde{E}_{k-3}\: + 2M b^{10}_{k-1} - 2 b^{10}_{k-2} \label{eq:b9_k_A} , \\
4M^2 k (k-4M i\omega) b^{10}_k & = & \tilde{H}_{k-1}\: b^{10}_{k-1} + \tilde{J}_{k-2}\: b^{10}_{k-2} + \tilde{E}_{k-3}\: b^{10}_{k-3} - 4M\lambda b^9_{k-1} - 2\lambda b^9_{k-2}\label{eq:b10_k_A} \\
\ea
where
\ba
\tilde{C}_k & = & 2M(k + 12 M i \omega k - 2k^2 + L - 4), \quad \tilde{H}_k =  2M(k + 12 M i \omega k - 2k^2 + L - 1), \nn \\
\tilde{D}_k & = & 4 + 12M i \omega k + L -k(k-1), \qquad \tilde{J}_k =  -2 + 12M i \omega k + L -k(k-1), \nn \\
\tilde{E}_k &=& 2 i \omega k . \nn
\ea
Now, we turn our attention to the BC for the even parity fields
$R^\pm_{1,3,5,6,7} $. We start with the recursion relations for
the outer BC for $R^\pm_1, R^\pm_3$ and $R^\pm_6 $:
\ba
2 i \omega k a^1_k & = & C^1_{k-1} a^1_{k-1} + (2-4Mi\omega) a^3_{k-1} + 2 a^5_{k-1} + 2 a^6_{k-1} + D^1_{k-2} a^1_{k-2} + D^3_{k-2} a^3_{k-2} - 12M a^5_{k-2}  \nn \\ & - &  20M a^6_{k-2} + E^1_{k-3} a^1_{k-3} + E^3_{k-3} a^3_{k-3} + 16 M^2 a^5_{k-3} + 56 M^2 a^6_{k-3} + F^3_{k-4}a^3_{k-4} - 48 M^3 a^6_{k-4},\nn \\ \label{eq:outBC1} \\
2 i \omega k a^3_k &=&  C^1_{k-1} a^3_{k-1} + 2 (a^1_{k-1} - a^5_{k-1} - a^6_{k-1} ) + D^1_{k-2} a^3_{k-2} \nn \\
& + & 4M (-a^1_{k-2} + a^5_{k-2} + 3 a^6_{k-2} ) + E^1_{k-3} a^3_{k-3} - 16M^2 a^6_{k-3}, \label{eq:outBC3} \\
2 i \omega k a^6_k &=&  C^1_{k-1} a^6_{k-1} + 2 (a^1_{k-1} - a^5_{k-1} - a^3_{k-1} ) + D^1_{k-2} a^6_{k-2} \nn \\
& + & 4M (-a^1_{k-2} + a^5_{k-2} + 3 a^3_{k-2} ) + E^1_{k-3} a^6_{k-3} - 16M^2 a^3_{k-3}, \label{eq:outBC6}
\ea
where
\ba
C^1_k & = & k(k+1) + 4M i \omega k - 2- L,\label{eq:rec_C1} \\
D^1_k & = & 2M (5+L-2k^2-3k), \quad D^3_k  =  2M (2k- 8 + 4M i \omega), \label{eq:rec_D1_D3} \\
E^1_k &= & 4 M^2 (k^2+2k-3), \quad E^3_k = 8 M^2 (5-2k), \quad F^3_k = 16 M^3 (k-2).\label{eq:rec_E1E3_F}
\ea
For the field $R^\pm_5$, we have:
\ba
2 i\omega k a^5_k & = & C^5_{k-1} a^5_{k-1} + 2L ( a^1_{k-1}-a^3_{k-1}-a^6_{k-1} ) + 2 a^7_{k-1} + D^5_{k-2} a^5_{k-2} - 10M a^7_{k-2} \label{eq:outBC5} \\
& + & 2ML (-2 a^1_{k-2} + 4 a^3_{k-2} + 5 a^6_{k-2} ) + E^5_{k-3} a^5_{k-3} + 4 M^2 (-2L a^3_{k-3} - 3L a^6_{k-3} + 3 a^7_{k-3} ) , \nn
\ea
where
\ba C^5_k & = & k(k+1) + 4M i \omega k - 4- L, \quad D^5_k = 2M(12-2k^2 - 3k + L), \label{eq:rec_C5} \\
E^5_k & = & 4M^2(k^2+2k-8). \label{eq:rec_E5}
\ea
And for $R^\pm_7$:
\ba
2 i \omega k a^7_k & = &C^7_{k-1}a^7_{k-1} + 2\lambda a^5_{k-1} + D^7_{k-2} a^7_{k-2} - 4M \lambda a^5_{k-2} + E^7_{k-3} a^7_{k-3}, \label{eq:outBC7} \ea
where
\ba
C^7_k & = & k(k+1) + 4M i \omega k - L +2, \quad D^7_k = 2M (L -3 -2k^2-3k), \label{eq:rec_C7} \\
E^7_k & = & 4 M^2 (k^2+2k+1). \label{eq:rec_E7}
\ea
Next, we present the recursion relations for the inner boundary conditions for the same fields in the same order. We start with:
\ba
8 M^3 k (4 M i \omega - k ) b^1_k & = & \t{C}^1_{k-1} b^1_{k-1} + \t{C}^3_{k-1} b^3_{k-1} - 8M^2 b^5_{k-1} + \t{D}^1_{k-2} b^1_{k-2} + \t{D}^3_{k-2} b^3_{k-2} \label{eq:inBC_1} \\
 & {} & -8M b^6_{k-2} + \t{E}^1_{k-3} b^1_{k-3} + 2(1+2 M i \omega) b^3_{k-3} + 2 b^5_{k-3} + 2 b^6_{k-3} - \t{F}^1_{k-4} b^1_{k-4}, \nn
\ea
where
\ba
\t{C}^1_k & = & 4 M^2 (1 + 3k^2 - L - 16 M i \omega k - k ), \quad \t{C}^3_k = 8 M^2 (2 M i \omega - k ), \label{eq:rec_tC} \\
\t{D}^1_k & = & 2M ( 3 k^2- 2k - 2L - 1 - 24 M i \omega k ), \ \ \ \t{D}^3_k = 4M ( 4 M i \omega - k -1 ), \label{eq:rec_tD} \\
\t{E}^1_k &= & k(k-1) - 16 M i \omega k - 2- L, \quad \t{F}^1_k = 2 i \omega k . \label{eq:rec_tEF}
\ea
\ba
4 M^2 k (4 M i \omega - k ) b^3_k & = & \t{G}^3_{k-1} b^3_{k-1} + 4M (b^1_{k-1} - b^5_{k-1} + b^6_{k-1} ) + \t{H}^3_{k-2} b^3_{k-2} \nn \\
& + & 2 ( b^1_{k-2} - b^5_{k-2} - b^6_{k-2} ) - \t{F}^1_{k-3} b^3_{k-3}, \label{eq:inBC_3}
\ea
\ba
4 M^2 k (4 M i \omega - k ) b^6_k & = & \t{G}^3_{k-1} b^6_{k-3} + 4M (b^1_{k-1} - b^5_{k-1} + b^3_{k-1} ) + \t{H}^3_{k-2} b^6_{k-2} \nn \\
& + & 2 ( b^1_{k-2} - b^5_{k-2} - b^3_{k-2} ) - \t{F}^1_{k-3} b^6_{k-3}, \label{eq:inBC_6}
\ea
where
\ba
\t{G}^3_k &= & 2M (2k^2- k - L +1 - 12 M i \omega k ), \label{eq:rec_G3} \\
\t{H}^3_k & =& k (k-1) - L -2 - 12 M i \omega k . \label{eq:rec_H3}
\ea
\ba
4 M^2 k (4 M i \omega - k ) b^5_k & = & \t{I}^5_{k-1} b^5_{k-1} + 2ML (2 b^1_{k-1} + b^6_{k-1} ) - 2M b^7_{k-1} + \t{J}^5_{k-2} b^5_{k-2} \nn \\
& + & 2L (b^1_{k-2} - b^3_{k-2} - b^6_{k-2} ) + 2 b^7_{k-2} - \t{F}^1_{k-3} b^5_{k-3}, \label{eq:inBC_5}
\ea
where
\ba \t{I}^5_k & = & 2M ( 2k^2- k - L + 4 - 12 M i \omega k ), \label{eq:rec_I5} \\
\t{J}^5_k & = & k(k-1) - L - 4- 12 M i \omega k . \label{eq:rec_J5}
\ea
And, finally
\ba
4 M^2 k (4 M i \omega - k ) b^7_k & = & \t{G}^3_{k-1} b^7_{k-1} + 4 M \lambda b^5_{k-1} + \t{K}^7_{k-2} b^7_{k-2}
  + 2 \lambda b^5_{k-2} \nn \\ &-&  \t{F}^1_{k-3} b^7_{k-3}, \label{eq:inBC_7}
\ea
where
\be \t{K}^7_k = k(k-1) - L +2 - 12 M i \omega k . \label{eq:rec_K7}\ee
\subsection{The Even Dipole ($\el=1, m=1 $) Mode}\label{sec:even_dipole_BC}
The recursion relations for $a^{1,3,6}_k, b^{1,3,6}_k$ do not change. However, we end up with new recursion relations for the inner and outer boundary conditions for $R^\pm_5 $
\ba
4 M^2 k (4 M i \omega - k ) b^5_k & = & \t{I}^5_{k-1} b^5_{k-1} + 2ML (2 b^1_{k-1} + b^6_{k-1} ) - 2M b^7_{k-1} + \t{J}^5_{k-2} b^5_{k-2} \nn \\
& + & 2L (b^1_{k-2} - b^3_{k-2} - b^6_{k-2} ) + 2 b^7_{k-2} - \t{F}^1_{k-3} b^5_{k-3}, \label{eq:inBC_5_1_1_A}
\ea
\ba
2 i\omega k a^5_k & = & C^5_{k-1} a^5_{k-1} + 2L ( a^1_{k-1}-a^3_{k-1}-a^6_{k-1} ) + 2 a^7_{k-1} + D^5_{k-2} a^5_{k-2} - 10M a^7_{k-2} \nn \\
& + & 2ML (-2 a^1_{k-2} + 4 a^3_{k-2} + 5 a^6_{k-2} ) + E^5_{k-3} a^5_{k-3}\nn\\ &+& 4 M^2 (-2L a^3_{k-3} - 3L a^6_{k-3} + 3 a^7_{k-3} ) \label{eq:outBC5_1_1_A} ,
\ea
where $\omega = m \Omega_0 $ and $ L \equiv \el(\el+1) $ as before.
The coefficients $ C^5_k, D^5_k, E^5_k, \t{F}^1_k, \t{I}^5_k, \t{J}^5_k $ are the same as before, displayed in Eqs. (\ref{eq:rec_C5}), (\ref{eq:rec_E5}), (\ref{eq:rec_tEF}), (\ref{eq:rec_I5}) and (\ref{eq:rec_J5}).

\subsection{The Static ($m=0$) Even Modes}\label{sec:even_stat_rec}
Recall that because we are now dealing with static modes, we no longer have in or outgoing waves at the boundaries. For the inner BC, we use the following ansatz:
\be R^-_i = \sum_{k=k_{\mrm{start}}}^\infty b^i_k (r-2M)^k \label{eq:IBC_even_stat_A} , \ee
where we now have three fields labelled by $i=1,3,5 $. New recursion relations for the inner BC are:
\ba
8M^3 k (k-2) b^1_ k & = & \b{F}^1_{k-1} b^1_{k-1} + \b{G}^1_{k-2} b^1_{k-2} + \b{G}^3_{k-2} b^3_{k-2} - 2M b^5_{k-2} \nn \\
& \quad & + \b{E}^1_{k-3} b^1_{k-3} + \b{E}^3_{k-3}b^3_{k-3} - b^5_{k-3},  \label{eq:rec_b1_stat_A} \\
4M k (k-1) b^5_k & = & \b{C}^5_{k-1} b^5_{k-1} - 4ML b^1_{k-1} + \b{D}^5_{k-2} b^5_{k-2} + 2L (b^3_{k-2} - b^1_{k-2} ),\nn \label{eq:rec_b5_stat_A} \\
\b{C}^3_{k-1} b^3_{k-1} & = & \b{C}^1_{k-1} b^1_{k-1} - 8 M^3 k b^1_k  + 4M^2 b^5_{k-1} + \b{D}^3_{k-2} b^3_{k-2} + \b{D}^1_{k-2} b^1_{k-2} \nn \\
& \quad & + 4M b^5_{k-2} + \b{E}^3_{k-3} b^3_{k-3} + \b{E}^1_{k-3} b^1_{k-3} + b^5_{k-3} , \label{eq:rec_b3_stat_A}
\ea
where
\ba
\b{C}^1_k & = & -4M^2 (k+1), \quad \b{C}^3_k  =  4M^2 k (k-1),\quad \b{C}^5_k = 2ML-4M(1+k^2), \label{eq:stat_C_A} \\
\b{D}^1_k & = & 2M(k-2), \quad \b{D}^3_k = 2M (L + k (1-2k)), \quad \b{D}^5_k = L-k(k+1), \label{eq:stat_D_A} \\
\b{E}^1_k &=& L +1 - k^2, \quad \b{E}^3_k = k-1, \quad \b{G}^3_k =  2Mk, \label{eq:stat_E_G_A}  \\
\b{F}^1_k & = & 4 M^2 (L+1+ 4k - 3k^2  ), \quad \b{G}^1_k = 2M (2L + 2 + 2k - 3k^2 ). \label{eq:stat_F_G_A}
\ea
For the outer boundary conditions, we make the following ansatz:
\be
R^+_i = \sum_{k=k_{\mrm{start}}}^\infty \f{a^i_k + \b{a}^i_k \ln{r}}{r^k} \label{eq:outBC_stat_even_A}.
\ee
Recall that the recursion relations for $a^1_k, a^3_k, a^5_k, \b{a}^1_k, \b{a}^3_k, \b{a}^5_k $ are determined by the three free parameters $ a^3_\el, a^5_\el, a^5_{\el+2} $. We now present these relations in their coupled form:
\ba
\h{C}^1_k a^1_k&=& (k+1) a^3_k + a^5_k - 2k \b{a}^1_k - \b{a}^3_k  \nn\\
& &-2M \left(\h{D}^1_{k-1} a^1_{k-1} + \h{D}^3_{k-1} a^3_{k-1} + a^5_{k-1} + \h{E}^1_{k-1} \b{a}^1_{k-1} - 2\b{a}^3_{k-1}\right) \nn \\
& & + 4M^2 \left(\h{F}^3_{k-2} a^3_{k-2} -\b{a}^3_{k-2} \right) \label{eq:outBC1_even_stat_A} , \\
\h{C}^1_k \b{a}^1_k & =&  (k+1) \b{a}^3_k + \b{a}^5_k -2M \left(\h{D}^1_{k-1} \b{a}^1_{k-1} + \h{D}^3_{k-1} \b{a}^3_{k-1} + \b{a}^5_{k-1} \right) \nn \\
& & +4M^2 \h{F}^3_{k-2} \b{a}^3_{k-2}, \label{eq:outBC_bar1_even_stat_A}
\ea
where
\ba
\h{C}^1_k & = & L+1-k^2, \quad \h{D}^1_k = k(k-1), \quad \h{D}^3_k = 2(k+1), \nn \\
\h{E}^1_k & = & 1-2k, \quad \h{F}^3_k = k+1, \nn
\ea
\ba
\h{C}^1_k a^3_k & = & (k+1) a^1_k - a^5_k -\b{a}^1_k - 2 k \b{a}^3_k \nn \\
& &-2M \left( \h{G}^3_{k-1} a^3_{k-1} + \h{G}^1_{k-1} a^1_{k-1} + \h{H}^3_{k-1} \b{a}^3_{k-1} - 2 \b{a}^1_{k-1} \right) \nn \\
& & + 4M^2 \left( \h{I}^3_{k-2} a^3_{k-2} + \h{J}^3_{k-2} \b{a}^3_{k-2} \right) \label{eq:outBC3_even_stat_A} , \\
\h{C}^1_k \b{a}^3_k & = & (k+1) \b{a}^1_k -\b{a}^5_k - 2M \left( \h{G}^3_{k-1} \b{a}^3_{k-1} + \h{G}^1_{k-1} \b{a}^1_{k-1} \right) \nn \\
& & + 4M^2 \h{I}^3_{k-2} \b{a}^3_{k-2} \label{eq:outBC_bar3_even_stat_A} ,
\ea
where
\ba
\h{G}^3_k & = & 2k^2-2-L, \quad \h{G}^1_k = 2k, \quad \h{H}^3_k = -4k, \nn \\
\h{I}^3_k & = & k^2-1, \quad \h{J}^3_k = -2k . \nn
\ea
\ba
\h{C}^5_k a^5_k & = & 2L (a^1_k - a^3_k) - \h{D}^5_k \b{a}^5_k + 2M \left( \h{E}^5_{k-1} a^5_{k-1} + 2L a^3_{k-1} + \h{D}^5_{k-1} \b{a}^5_{k-1} \right) \label{eq:outBC5_even_stat_A} , \\
\h{C}^5_k \b{a}^5_k & = & 2L (\b{a}^1_k -\b{a}^3_k ) + 2M \left(\h{E}^5_{k-1} \b{a}^5_{k-1} + 2L \b{a}^3_{k-1} \right), \label{eq:outBC_bar5_even_stat_A}
\ea
where
\be \h{C}^5_k  =  L + k(1-k), \quad \h{D}^5_k = 2k-1, \quad
\h{E}^5_k  =  k(1-k) + 2 . \nn \ee

\end{document}